\documentclass[11pt]{article}
\usepackage{graphicx}
\usepackage{epsfig}
\newcommand{\ja}{{j_a}}
\newcommand{\jb}{{j_b}}

\newcommand{\la}{{l_a}}
\newcommand{\lb}{{l_b}}

\newcommand{\beq} {\begin{equation}}
\newcommand{\eeq} {\end{equation}}
\newcommand{\what}[1]{\widehat #1}

\newcommand{\ji} {j_i}
\newcommand{\mi} {m_i}

\newcommand{\pon} {p_1}
\newcommand{\ptw} {p_2}
\newcommand{\hon} {h_1}
\newcommand{\htw} {h_2}
\newcommand{\ron} {r_1}
\newcommand{\rtw} {r_2}

\newcommand{\bsigma}{\mbox{\boldmath $\sigma$}}

\newcommand{\bkey}{{\bf k}}
\newcommand{\bpi}{{\bf p}}
\newcommand{\bqu}{{\bf q}}
\newcommand{\br}{{\bf r}}
\newcommand{\half}{\frac{1}{2}}
\newcommand{\e}[1]{ {\rm e}^{#1} }
\newcommand{\threej}[6]{ \left( \begin{array}{ccc}
                               #1 & #2 & #3 \\
                               #4 & #5 & #6 
                             \end{array}
                        \right) }

\begin{document}

\begin{titlepage}
\thispagestyle{empty}
\begin{center}
{\Large \bf A model for two-proton emission induced by electron scattering} 

\vspace{1.5cm}
{\large Marta Anguiano and Giampaolo Co'} \\
\vspace{.3cm}
{Dipartimento di Fisica, Universit\`a di Lecce and
I.N.F.N. sez. di Lecce, I-73100 Lecce, Italy}

\vskip 1cm

{\large Antonio M. Lallena}\\
\vspace{.3cm}
{Departamento de F\'{\i}sica Moderna, Universidad de Granada,
E-18071 Granada, Spain} 
\end{center}

\vskip 1.5 cm 

\begin{abstract}
A model to study two-proton emission processes induced by electron
scattering is developed.  The process is induced by one-body
electromagnetic operators acting together with short-range
correlations, and by two-body $\Delta$ currents.  The model includes
all the diagrams containing a single correlation function. A test of
the sensitivity of the model to the various theoretical inputs is
done. An investigation of the relevance of the $\Delta$ currents is
done by changing the final state angular momentum, excitation energy
and momentum transfer. The sensitivity of the cross section to the
details of the correlation function is studied by using realistic and
schematic correlations. Results for $^{12}$C, $^{16}$O and $^{40}$Ca
nuclei are presented.
\end{abstract} 

\vskip 1cm
PACS number(s): 21.10.Ft, 21.60.-n
\end{titlepage}

\section{INTRODUCTION}
\label{sec:intro}

This article belongs to a series dedicated to the study of the
effects of the short range correlations (SRC) in processes induced by
electromagnetic probes on atomic nuclei \cite{co01}-\cite{ang02}.  

The SRC are produced by the strong repulsion of the bare
nucleon-nucleon potential at internucleon distances smaller than
$\sim$ 0.5 fm. Every nuclear structure calculation that uses bare
nucleon-nucleon potentials requires the presence of SRC, however, in
medium-heavy nuclei, phenomena which can be unambiguously attributed
to them have not yet been identified.

A well known effect of the SRC is the depletion of the occupation
probability of the quasi-hole states. Unfortunately, an analogous
effect is also produced by the coupling of the single particle wave
function with the low-lying phonons generated by collective nuclear
vibrations. The inclusion of both effects seems to be necessary to
account for the empirical occupation numbers \cite{ang01,fab01}.

A more striking, and even better known, effect of the SRC is the
various order of magnitude enhancement of the nucleon momentum
distribution $n(k)$ at high momentum values \cite{ant88}.
Unfortunately, this quantity is not directly observable.  It is
therefore necessary to find measurable quantities related to it,
hoping that they are sufficiently sensitive to $n(k)$ to allow for an
unambiguous identification of the SRC effects.

In these last years we have developed a model to describe the
responses of finite nuclear systems within the framework of the
correlated basis function theory. The model is inspired to the nuclear
matter works of Refs. \cite{fan87,fab89}.  With respect to these
calculations, in our model the cluster expansion is truncated and only
those terms containing a single correlation function are considered.
The set of diagrams taken into account conserves the proper
normalization of the many-body wave function.

We tested the validity of this treatment by comparing the nuclear
matter charge responses calculated with our model with those obtained
by the full expansion \cite{ama98}. The excellent agreement between
these two results gave us confidence in extending the application of
the model to other cases.  Since our description of the nuclear
excitations does not treat properly collective states, we have selected
situations dominated by single particle dynamics.

With our model we have studied the electromagnetic form factors of
discrete states with large angular momentum \cite{mok00}.  These
states are dominated by a single, or at most a few, particle-hole
excitation.  We found that the SRC produce small effects and are
unable to account for the quenching required to reproduce the data.
Electron scattering inclusive reactions in the quasi-elastic region
have been investigated in \cite{co01}. In this case the effect of the
correlations is much smaller than that produced by the final state
interaction, and therefore difficult to disentangle. We have also
investigated phenomena where a single nucleon is emitted by the
interaction with a virtual \cite{mok01} or real \cite{ang02} photon.
The study of (e,e'p) reactions show a very small sensitivity to the
SRC. The dominant effect beyond the independent particle model is
produced by the final state interaction.  Nucleon emission induced by
real photons shows a larger sensitivity to SRC, especially for
excitation energies around 200 MeV and large nucleon emission angles
\cite{ang02}.  Unfortunately, in this case, the meson exchange
currents produce effects even larger than those induced by the SRC.

The difficulty in identifying clear SRC effects in our previous
studies was due to the presence of the large contribution of the
uncorrelated one-body responses that dominates the cross sections.  A
possibility of eliminating the one-body responses is offered by
two-nucleon emission experiments. Only two-body transition operators
can induce the emission of two nucleons. These operators can
effectively be constructed by one-body operators acting on a correlated
nucleonic pair, or by meson exchange currents.  If the two emitted
nucleons are of the same type, two protons for example, the meson
exchange currents contributions produced by the exchange of a single
charged mesons do not contribute. These facts make two-proton emission
experiments an ideal tool to investigate SRC.

From the theoretical point of view, two-nucleon emission processes have
been systematically studied in these years by the Pavia and Gent
groups.  The approach of the Pavia group \cite{giu91,giu97} is based
on the two-nucleon spectral function, and, from the theoretical point
of view, it is a straightforward extension of the well tested approach
used to describe single nucleon emission processes.  Recently
two-nucleon spectral functions calculated with microscopic theories
\cite{geu96,giu98} have been used.
The Gent group \cite{ryc95}-\cite{ryc97}, calculates transition
amplitudes produced by effective two-body operators that are formed by
connecting the traditional one-body electromagnetic operators to
two-nucleon correlation function.

The experimental situation is quite promising. After testing the
feasibility of the experiment \cite{kes95,zon95}, angular correlations
of the cross sections have been measured \cite{ond97,sta00}. Other
experiments of this kind have been planned.  

In this article we present the results obtained by applying our model
to the study of two-proton emission induced by electron scattering.  A
comparison of our results with the available data is out the scope of
the present article.  Our aim is to investigate the sensitivity of the
(e,e'2p) cross section to the details of the SRC. To this purpose we
have studied the influence on the cross section of the various
theoretical inputs required by the calculation.

A summary of the basic formulae used to describe (e,e'2p) processes
and electromagnetic currents is given in Section \ref{sec:cs}. In
Section \ref{sec:nucmod} we describe our nuclear model and in Section
\ref{sec:app} we discuss the results we have obtained in $^{12}$C,
$^{16}$O and $^{40}$Ca target nuclei. In the last section we draw our
conclusions.

\section{THE CROSS SECTION}
\label{sec:cs}
A detailed derivation of the cross section for double
coincidence experiments can be found in \cite{bof96}.  In this section
we briefly recall the expressions used in our calculations. We work in
natural units ($\hbar=c=1, e^2=1/137.04$) and employ the conventions
of Bjorken and Drell \cite{bjo64}.  The initial and final electron
four-vectors are respectively $k \equiv (\epsilon,\bkey)$ and $k'
\equiv (\epsilon',\bkey')$, and $q \equiv (\omega,\bqu)= k - k'$ is
the four-momentum transfer. The four-momenta of the emitted nucleons
are indicated with $\pon \equiv (\epsilon_1,\bpi_1)$ and $\ptw \equiv
(\epsilon_2,\bpi_2)$.

\begin{figure}[hb]
\begin{center}
\vspace*{0.cm}
\hspace*{-0.0 cm}
\leavevmode
\epsfysize = 280pt
\epsfbox[70 200 500 650]{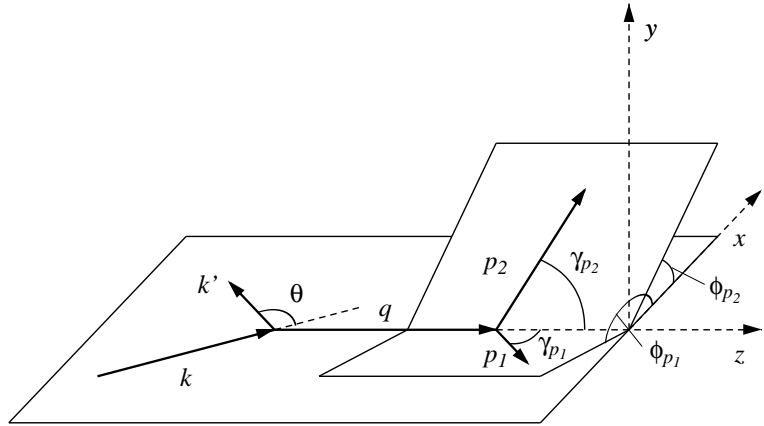}
\end{center}
\vspace*{-4.5cm}
\caption{\small Reference system used in our calculations.
}
\label{fig:axis}
\end{figure}

The reference system we adopt is shown in Fig. \ref{fig:axis}. The
scattering plane is defined by the electron vectors $\bkey$ and
$\bkey'$, $\theta$ is the angle of the scattered electron and the
quantization axis is taken along the direction of $\bqu$.  The
directions of the emitted nucleons are determined by the angles
$\gamma_{p_i}$ and $\phi_{p_i}$, $i=1,2$.  To simplify the formalism our
calculations have been done considering two different angles
$\varphi_{p_i}$ and $\theta_{p_i}$.  The relation to the angles described in
the figure is
\[
\varphi_{p_i} = \left\{
\begin{array}{ll}
\phi_{p_i}  & \mbox{if $0 \leq \phi_{p_i} \leq \pi$} \\
\phi_{p_i} - \pi  & \mbox{if $\pi \leq \phi_{p_i} \leq 2\pi$}   
\,\,\, ,
\end{array}
\right.
\]
and
\[
\theta_{\pon} = \left\{
\begin{array}{ll}
2 \pi - \gamma_{\pon}  & \mbox{if $0 \leq \gamma_{\pon} \leq \pi$} \\
\gamma_{\pon}  & \mbox{if $\pi \leq \gamma_{\pon} \leq 2 \pi$}   
\,\,\, ,
\end{array}
\right.
\]
\[
\theta_{\ptw} = \left\{
\begin{array}{ll}
\gamma_{\ptw}  & \mbox{if $0 \leq \gamma_{\ptw} \leq \pi$} \\
2 \pi- \gamma_{\ptw}  & \mbox{if $\pi \leq \gamma_{\ptw} \leq 2\pi$}   
\,\,\, .
\end{array}
\right.
\]
Note that the $\gamma_{\ptw}$ angle in Fig. \ref{fig:axis} has been
defined positive in the half plane containing the scattered
electron. The convention for  $\gamma_{\pon}$ is opposite.

We have derived the cross section expression by considering that the
electron wave functions are plane wave solutions of the Dirac
equation. We suppose that only one photon is exchanged between the
electron and the nucleus and we neglect all the terms depending on the
electron rest mass.  With these approximations we obtain
\label{sec:xsect}
\beq
\frac{{\rm d}^8 \sigma}
{{\rm d}\epsilon' {\rm d}\Omega_e {\rm d}\epsilon_1 {\rm d}\Omega_{\pon}
{\rm d}\Omega_{\ptw} } =
\frac{K}{(2 \pi)^6} \, \sigma_{\rm Mott} \, f_{\rm rec} \,
\left(  v_l w_l + v_t w_t + v_{tl} w_{tl} + v_{tt} w_{tt}   \right)
\, ,
\label{eq:cross}
\eeq
where we have indicated with $\sigma_{\rm Mott}$ the Mott cross section
\begin{equation}
\label{eq:mott}
\sigma_{\rm Mott} \, = \,
\left(\frac {e^2 \cos(\theta/2)}{2\epsilon \sin^2(\theta/2)} \right)^2
\, ,
\end{equation}
and with $f_{{\rm rec}}$ the recoil factor
\beq
\label{eq:rec}
f_{{\rm rec}}^{-1}= 1+ \frac{m}{M_{A-2}}
\left[ 1+ \frac{|\bpi_1|}{|\bpi_2|}\cos \theta_{12} - 
          \frac{|\bqu|}{|\bpi_2|}\cos \theta_{\ptw}  \right]
\, ,
\eeq
where $\theta_{12}$ is the angle between the momenta of the 
two emitted nucleons, $\bpi_1$ and $\bpi_2$, and $m$ and $M_{A-2}$ are
the nucleon and rest nucleus masses, respectively.
The factor $K$ is
\beq
K=m^2 \, |\bpi_1| \, |\bpi_2|
\, ,
\eeq
because we used non relativistic kinematics to describe the motion of
the two emitted nucleons.  The expression (\ref{eq:cross}) has been
obtained by integrating on the energy $\epsilon_2$ of one of the
emitted particles and using the energy conservation.

In the plane wave approximation the leptonic and hadronic vertexes can
be factorized. In our expression (\ref{eq:cross}) the terms $v$ come
from the leptonic vertex and depend only from kinematic variables
\begin{eqnarray}
\label{eq:vl}
v_l &=& \left(\frac{q_\mu^2}{\bqu^2} \right)^2 \, ,\\
\label{eq:vt}
v_t &=& \tan^2 \frac{\theta}{2} - \half \frac{q_\mu^2}{\bqu^2} \, ,\\
\label{eq:vtl}
v_{tl} &=& \frac {q_\mu^2}{\sqrt{2}\bqu^2} \,
 \left( \tan^2 \frac{\theta}{2} - 
\frac{q_\mu^2}{\bqu^2} \right)^\half \, ,\\
\label{eq:vtt}
v_{tt} &=& \half \frac{q_\mu^2}{\bqu^2} \, .
\end{eqnarray}

The information about the nuclear structure is included in the $w$
factors. Because of the current conservation only three components of
the four-vector current are independent.  We choose the charge
$\rho(\bqu)$ and the two transverse components in spherical
coordinates
\beq
J_{\pm}=\mp \frac{1}{\sqrt{2}} 
\left(J_x \pm i J_y  \right) \, .
\label{eq:jpm}
\eeq

In analogy to the (e,e'p) case \cite{mok01} the $w$ factors can be
expressed as
\begin{eqnarray}
\label{eq:wl}
w_l &=& \langle \Psi_i | \rho^\dagger (\bqu) | \Psi_f \rangle 
        \langle \Psi_f | \rho (\bqu) |\Psi_i \rangle  \, , \\ 
\label{eq:wt}  
w_t &=& \langle \Psi_i | J_{-}^\dagger (\bqu) | \Psi_f \rangle 
        \langle \Psi_f | J_{-} (\bqu) |\Psi_i \rangle +
        \langle \Psi_i | J_{+}^\dagger (\bqu) | \Psi_f \rangle 
        \langle \Psi_f | J_{+} (\bqu) |\Psi_i \rangle \, ,\\  
\label{eq:wtl}
w_{tl} &=& 2 Re  \left( \langle \Psi_i | \rho^\dagger (\bqu) | \Psi_f \rangle 
           \langle \Psi_f |  J_{-} (\bqu) |\Psi_i \rangle
        - \langle \Psi_i | \rho^\dagger (\bqu) | \Psi_f \rangle 
          \langle \Psi_f | J_{+} (\bqu) | \Psi_i \rangle 
           \right) \, ,\\
\label{eq:wtt}   
w_{tt} &=&  2 Re  \left(
           \langle \Psi_i | J^\dagger_{+} (\bqu) | \Psi_f \rangle 
           \langle \Psi_f |  J_{-} (\bqu) |\Psi_i \rangle
           \right) \, ,
\end{eqnarray}
where we have indicated with $|\Psi_i\rangle$ and $|\Psi_f\rangle$
the initial and final states of the full hadronic system.  We do not
consider the polarization of the emitted nucleons, therefore, in the
previous equations, a sum on the third components of the spin of the
emitted particles and of the angular momentum of the residual
nucleus is understood.

The one-body electromagnetic operators we have used are produced by the
charge operator
\begin{equation}
\label{eq:charge1}
\rho({\bf r}) \, = \, 
\sum^A_{k=1} \frac{1+\tau_3(k)}{2} \, 
\delta({\bf r}-{\bf r}_k) \, ,
\end{equation}
and by the magnetization current operator
\begin{equation}
\label{eq:mag}
J^{\rm M}({\bf r}) \, = \, 
\sum^A_{k=1} \frac{1}{2m_k} \,
\left(\mu^{\rm P}\frac{1+\tau_3(k)}{2} + 
\mu^{\rm N}\frac{1-\tau_3(k)}{2} \right)
\nabla \times \delta({\bf r}-{\bf r}_k) \, \bsigma(k) \, .
\end{equation}
In the previous equations $m_k$ indicates the rest mass of $k$--th
nucleon, $\mu^{\rm P}$ and $\mu^{\rm N}$ the anomalous magnetic moment
of the proton and the neutron, respectively, $\bsigma(k)$ the Pauli spin
matrix of the $k$-th nucleon and $\tau_3(k)=1$ ($-1$) if the $k$-th nucleon is a proton
(neutron). In our calculations the nucleonic internal structure has
been considered by folding the point-like responses with the
electromagnetic nucleon form factors of Ref. \cite{hoe76}.
To simplify the calculations we did not include the convection
current because we know that its contribution is small in the
quasi-elastic region \cite{co01,ang02}.

\begin{figure}[hb]
\includegraphics[bb=250 650 480 262, angle=90,scale=1.0]
       {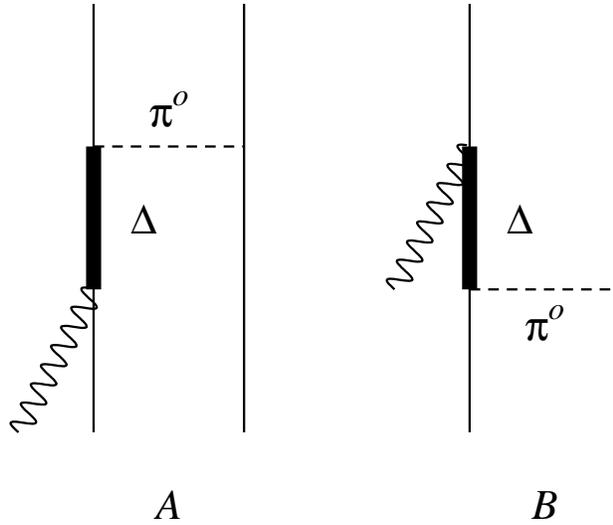}
\vspace*{-1.cm}
\caption{\small Meson exchange currents diagrams considered in our
       calculations. The wiggly lines represent the virtual photon
       exchanged with the electron, the full thin lines the
       nucleons. Since we describe the emission of two protons, the
       exchanged pion is chargeless.
}
\label{fig:delta}
\end{figure}

Since we have considered only the emission of two like particles (two
protons), the two-body currents induced by the exchange of charged
mesons do not contribute. The main two-body current contribution
arises from diagrams like those of Fig. \ref{fig:delta} where the
exchanged $\pi$ meson is chargeless.  These diagrams correspond to the
$\Delta$-isobar currents for which we have used the following
expression
\begin{eqnarray} 
\nonumber
{\bf j}^{\Delta}(\br) \, = &&
\frac{ f_{\pi N \Delta } f_{\pi N N} f_{\gamma N \Delta} } { 9 m^3_{\pi}} \,
\, \sum_{k<l=1}^A \, \tau_3(l) 
\, \nabla \times \delta(\br - \br_k) \\
\nonumber
&&
\left[ i\, 2 \, G^-_\Delta \, \bsigma(l) \times \nabla \,
 - \, 4 \, G^+_\Delta \,\nabla \right]  
\left[ \bsigma(k) \cdot \nabla \,  
{\cal H}(\br-\br_l,\varepsilon_l) \right]
\\
& &
+ (k \longleftrightarrow l)
\, ,
\label{eq:delta}
\end{eqnarray}
where the factors $f_{\pi N \Delta }$, $f_{\pi N N}$ and
$f_{\gamma N \Delta}$ are the coupling constants related to the
processes indicated by the lower labels, and $m_\pi$ is the pion mass.
The Fourier transform of the above expression has been multiplied by
the $\Delta$ electromagnetic form factor $F_{\Delta}$ 
that we have considered in its dipole form
\beq
F_{\Delta} (|\bqu|,\omega) = 
\left[
1 - \frac {\omega^2 - \bqu^2}{\Lambda^2}
\right]^{-2}
\,\,\, ,
\eeq 
with $\Lambda$=855 MeV when $\omega$ and $\bqu$ are expressed in MeV.
The two coupling constants $G^+_\Delta$ and $G^-_\Delta$ are
related to the two diagrams shown in Fig. \ref{fig:delta} and
are defined as
\beq
G^\pm_\Delta = G_A \pm G_B \, ,
\eeq
Following the indications of Refs. \cite{wil96,wil97}, we have taken
the coupling
\beq
G_A = \left[m_\Delta - \sqrt{s} - \frac{i}{2}
  \Gamma_\Delta(\sqrt{s})
\right]^{-1}
\label{eq:gi}
\, ,
\eeq
where
\beq
s = 2 \omega (m_n + \epsilon_h) +
             (m_n + \epsilon_h)^2
\, ,
\eeq
and $\epsilon_h$ is energy of the hole single particle level.  The
expression of the imaginary part is taken from Ref. \cite{ose82},
\beq
\Gamma_\Delta (\sqrt{s}) = \frac 
   { 8 \,\, f_{\pi N N} \,\, (m_\Delta - m_n) \,\, (s - m_\pi^2)^{3/2} }
   { 3 \,\, m_\pi^2 \,\, \sqrt{s}  } \, .
\eeq
On the other hand,
\beq
G_B = (m_\Delta - m_n)^{-1}
\label{eq:gii} 
\, .
\eeq
In the previous equations $m_\Delta=1232$ MeV and 
$m_n=938.9$ MeV are the $\Delta$ and nucleon masses, respectively.
Finally, with the function ${\cal H}(\br)$ we indicated the Fourier
transform of the dynamical pion propagator
\begin{equation}
{\cal H}(\br-\br_l,\varepsilon_l)
=\int \frac{{\rm d}^3k}{(2\pi)^3}\,
\frac{F_{\pi\rm N}(k,\varepsilon_l)\, \e{i\bkey\cdot(\br-\br_l)}}%
{k^2+m_{\pi}^2-\varepsilon_l^2} \,
\, ,
\label{eq:hache}
\end{equation}
where $F_{\pi\rm N}$ is the pion-nucleon form factor and 
$\varepsilon_l$ is the energy of the exchanged pion
obtained as the difference between the energies of the final and
initial states of the $l$-th nucleon.
To simplify our calculations we set the pion-nucleon form factor
$F_{\pi\rm N}(k,\varepsilon)=1$. We have verified that it is a very
good assumption in the energy and momentum region we are interested to
explore \cite{ama93}.

\section{THE NUCLEAR MODEL}
\label{sec:nucmod}
In the previous section we have presented the expressions of the cross
sections and of the electromagnetic operators used in our
calculations. Now we have to specify how we describe the initial and
final nuclear states to be used to calculate the transition matrix
elements in Eqs. (\ref{eq:wl})-(\ref{eq:wtt}).  To simplify the
presentation we will be first concerned with the description of the
final state which, asymptotically, has two nucleons in the continuum
and two holes. In a following step we shall discuss how we describe
the correlated many-body states. Our numerical calculations have been
restricted to the case of the emission of two protons, but the
formalism is general enough to deal with the emission of two generic
nucleons.

We describe the single particle wave function of a nucleon moving in
the continuum with momentum $\bpi$ and with third components of spin
and isospin $s$ and $t$, respectively, as
\beq
\nonumber
\psi(\bpi \cdot \br) \, \chi_s \, \chi_t \, = \,
\frac{4 \pi}{|{\bf p}|} \, \sum_{l=0}^\infty 
\sum_{\mu=-l}^{l} \sum_{j=l \pm 1/2}
(-i)^{\,l} \, R^t_{l,j}(pr) \, Y^*_{l \mu}(\Omega_p) \,
\langle l \mu \half s | j m \rangle \, Y^m_{l,j}(\Omega_r) \, \chi_t
\,\,\, .
\label{eq:spwf}
\eeq
In the above equation we have done the multipole decomposition of the
wave function and we have indicated with 
$ \langle l \mu \half s | j m \rangle$
the Clebsch-Gordan
coefficients, with $ Y_{l \mu} $ the spherical harmonics and with
\begin{equation}
Y^m_{l,j}(\Omega) \, = \, 
\sum_{\mu s} \, \langle l \mu \half s | j m \rangle \, Y_{l \mu}(\Omega)
\, \chi_s \, \equiv \, |l\, \half\, j \, m \rangle
\, ,
\end{equation}
the spin spherical harmonics \cite{edm57}. The symbols
$\chi_s$ and $\chi_t$ indicate, respectively, the spin and isospin 
parts of the wave function and $R^t_{l,j}(pr)$ its radial part, obtained
by solving the Schr\"odinger equation with a spherical mean field
potential. 

The many-body final state is characterized by the momenta of the
emitted nucleons $\bpi_1$ and $\bpi_2$, their spin and isospin
components $s_1$, $s_2$, $t_1$ and $t_2$, the total angular momentum
of the residual nucleus $J_h$ and its z-axis component $M_h$, and the
orbital and total angular momenta, $l_{\hon}$, $j_{\hon}$, $l_{\htw}$
and $j_{\htw}$, of the single particle states where the nucleons are
coming from
\begin{equation}
\hspace*{-.5cm}
| \Psi_{\rm f}(\bpi_1,s_1,t_1,\bpi_2,s_2,t_2;J_h,M_h;
       l_{\hon},j_{\hon},l_{\htw},j_{\htw})\, \rangle
\, = \,
a^+_{\pon s_{1}} a^+_{\ptw s_{2}} 
\left[ \tilde{a}_{\,l_{\hon} , j_{\hon}} 
\otimes \tilde{a}_{\,l_{\htw},j_{\htw}}
\right]^{J_{h}}_{M_{h}}
| \Psi(0,0) \,  \rangle 
\, .
\end{equation}
In the above equation $a^+_{p s}$ creates a particle with momentum
$|\bpi|$ and spin third component $s$ (the isospin indexes have been
dropped to simplify the writing), and we have defined the annihilation
operator as
\beq
\tilde{a}_{\,l,j,m} = (-1)^{j+m} a_{\,l,j,-m}
\,\,\, ,
\eeq
in order to deal with irreducible spherical tensors. The symbol
$\otimes$ indicates the tensor product between two of such tensors
\cite{edm57}. 

The nuclear ground state is represented
by $| \Psi(0,0) \, \rangle$ defined as
\beq
| \Psi(0,0) \rangle = 
\frac { F |\Phi (0,0) \rangle }
{ \langle \Phi (0,0) | F^+  F | \Phi (0,0) \rangle }
\, ,
\eeq
where $F$ is the correlation function and $ |\Phi (0,0) \rangle $
is the Slater determinant describing a mean field ground state.
We work with doubly magic nuclei, therefore the two zeros on the
ground state symbol indicate the angular momentum value and its
projection on the quantization axis. 

By considering the multipole expansion of
the single particle continuum wave function we can write
\begin{eqnarray}
\nonumber
\hspace*{.5cm}
| \Psi_{\rm f}(\bpi_1,s_1,t_1,\bpi_2,s_2,t_2;J_h,M_h;
               l_{\hon},j_{\hon},l_{\htw},j_{\htw})\, \rangle &&\\
&&
\nonumber
\hspace*{-4.8cm}= \frac{(4 \pi)^2} {|{\bpi_1}|\,|{\bpi_2}|} \,
\sum_{l_{\pon} \mu_{\pon}} \sum_{j_{\pon} m_{\pon}} 
\sum_{l_{\ptw} \mu_{\ptw}} \sum_{j_{\ptw} m_{\ptw}} \,
 (-i)^{l_{\pon}+l_{\ptw}} \\
&~&        
\nonumber
\hspace*{-4.5cm}Y_{l_{\pon} m_{\pon}}(\Omega_{\pon}) \,
         Y_{l_{\ptw} m_{\ptw}}(\Omega_{\ptw}) \,
 \langle l_{\pon} \mu_{\pon} \half s_1 | j_{\pon} m_{\pon} \rangle \,
 \langle l_{\ptw} \mu_{\ptw} \half s_2 | j_{\ptw} m_{\ptw} \rangle \\
&~&
\hspace*{-4.5cm}
| \Psi_{\rm f} ( l_{\pon}, \mu_{\pon}, j_{\pon}, m_{\pon};
         l_{\ptw}, \mu_{\ptw}, j_{\ptw}, m_{\ptw};
         l_{\hon},   j_{\hon}; l_{\htw}, j_{\htw}; J_h,M_h ) \rangle        
\,\,\, .
\label{eq:muexp}
\end{eqnarray}

The many-body final state on the right hand side of the above equation
is described as
\begin{eqnarray}
\nonumber
\hspace*{.5cm}
| \Psi_{\rm f} ( l_{\pon}, \mu_{\pon}, j_{\pon}, m_{\pon};
         l_{\ptw}, \mu_{\ptw}, j_{\ptw}, m_{\ptw};
         l_{\hon},   j_{\hon}; l_{\htw}, j_{\htw}; J_h,M_h ) \rangle &&\\
&& \hspace{-8.5cm}
= \, \sum_{m_{\hon} m_{\htw}} \, (-1)^{ j_{\hon}+m_{\hon}+j_{\htw}+m_{\htw} } \,
\langle j_{\hon}\,-m_{\hon}\,j_{\htw}\,-m_{\htw} | J_h M_h \rangle \,
\frac{F |\Phi_{\rm f} \rangle}
{ \langle \Phi_{\rm f} | F^+ F |\Phi_{\rm f} \rangle ^\half }
\, .
\end{eqnarray}
The Slater determinant $|\Phi_{\rm f} \rangle$ is built on the ground state
Slater determinant $|\Phi(0,0) \rangle$ by substituting the hole wave
functions $\hon$ and $\htw$ with the particle wave functions $\pon$
and $\ptw$. In this specific case, the various single particle wave
functions are not coupled to a total angular momentum, therefore the
quantum numbers characterizing $|\Phi_{\rm f} \rangle$ are the orbital, $l$,
and total, $j$, angular momenta of the single particle levels, their
$z$-axis projections, $\mu$ and $m$, respectively, and the third
components of the isospin.

The evaluation of the $w$ functions in Eqs.
(\ref{eq:wl})-(\ref{eq:wtt}) requires the calculation of transition
matrix elements of the type
\begin{eqnarray}
\nonumber
\langle \Psi_{\rm f} | O_\eta(\bqu) |\Psi (0,0)\rangle &=&
\frac{(4 \pi)^2} {|{\bpi_1}|\,|{\bpi_2}|} \,
\sum_{l_{\pon} \mu_{\pon}} \sum_{j_{\pon} m_{\pon}} 
\sum_{l_{\ptw} \mu_{\ptw}} \sum_{j_{\ptw} m_{\ptw}} \sum_{m_{\hon} m_{\htw}} \,
 i^{\, l_{\pon}+l_{\ptw}} \, (-1)^{ j_{\hon}+m_{\hon}+j_{\htw}+m_{\htw} }
\\
&~&        
\nonumber
 \langle l_{\pon} \mu_{\pon} \half s_1 | j_{\pon} m_{\pon} \rangle \,
 \langle l_{\ptw} \mu_{\ptw} \half s_2 | j_{\ptw} m_{\ptw} \rangle \,
 \langle j_{\hon}\,-m_{\hon}\,j_{\htw}\,-m_{\htw} | J_h M_h \rangle
\\
&~&
 Y^*_{l_{\pon} m_{\pon}}(\Omega_{\pon}) \,
 Y^*_{l_{\ptw} m_{\ptw}}(\Omega_{\ptw}) \, 
 \xi[ O_\eta(\bqu):\pon,\ptw,\hon,\htw]
\, ,
\label{eq:psif2}
\end{eqnarray}
where 
\beq
O_{\eta}(\bqu)\, =
\, \int {\rm d}^3 r \, e^{- i \bqu \cdot \br } \, O_{\eta}(\br)
\, ,
\label{eq:opft}
\eeq
indicates a one-body electromagnetic operator and we have defined
\begin{eqnarray}
\nonumber
\xi[ O_\eta(\bqu):\pon,\ptw,\hon,\htw] &=&
\frac{ \langle \Phi_{\rm f} | F^+  O_\eta(\bqu) F | \Phi(0,0) \rangle }
{
\langle \Phi_{\rm f} | F^+ F | \Phi_{\rm f} \rangle ^\half
\langle \Phi(0,0) | F^+ F | \Phi(0,0) \rangle ^\half} \\
&=&
\frac{ \langle \Phi_{\rm f} | F^+  O_\eta(\bqu) F | \Phi(0,0) \rangle }
{\langle \Phi_{\rm f} | F^+ F | \Phi_{\rm f} \rangle }
\left[
\frac{ \langle \Phi_{\rm f} | F^+ F | \Phi_{\rm f} \rangle }
{ \langle \Phi(0,0) | F^+ F | \Phi(0,0) \rangle }
\right]^\half
\,\,\, .
\label{eq:xi}
\end{eqnarray}

In our calculations we have used purely scalar correlations defined
as:
\beq
F(1,2,...A)=\prod^A_{i<j} f(r_{ij}) 
\,\,\, ,
\eeq
where $r_{ij}=|{\bf r}_i-{\bf r}_j|$ is the distance between the
positions of the particles $i$ and $j$.

The two factors in Eq. (\ref{eq:xi}) are separately evaluated by
expanding numerator and denominator in powers of the two-body
short-range correlation function $f$. The presence of the denominator
and the energy conservation eliminate the unlinked diagrams.

Since in our calculations the correlation functions are purely scalar
they commute with the operator $O_\eta (\bqu)$ therefore:
\begin{eqnarray}
\nonumber
\xi[ O_\eta(\bqu):\pon,\ptw,\hon,\htw] &=&
\langle \Phi; \pon,\ptw,\hon,\htw | 
O_{\eta} (\bqu) F^2 |\Phi; 0 0, +1 \rangle_L \\
&=&  \langle\Phi; \pon,\ptw,\hon,\htw |
O_{\eta} (\bqu) \prod^A_{i<j}(1+h_{ij})|\Phi(0,0) \rangle_L \,
\,\,\, ,
\label{eq:xi3}
\end{eqnarray}
where we have used the function $h_{ij}=f^2(r_{ij})-1$ and
the subindex $L$ indicates that only the linked diagrams are
considered. 

\begin{figure}[ht]
\begin{center}
\vspace*{3.cm}
\hspace*{-0.0 cm}
\leavevmode
\epsfysize = 300pt
\epsfbox[70 200 500 650]{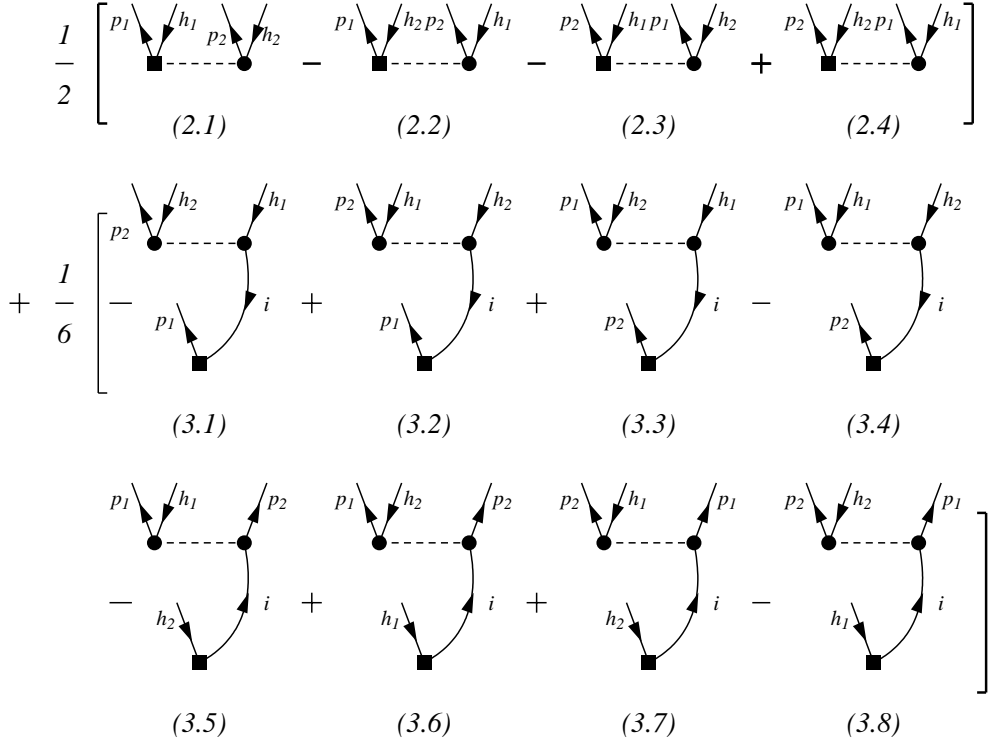}
\end{center}
\vspace*{-3.5cm}
\caption{\small Meyer-like diagrams representing the one-body currents
       matrix elements included in our calculations. The dashed lines
       represent the function $h(r)=f^2(r)-1$, where $f(r)$ is the
       correlation function. The black squares are the points where
       the electromagnetic one-body operator ${\cal O}_{\eta}(\br)$ is
       acting. The oriented full lines represent single particle wave
       functions and one has to understand a sum on all the hole
       states labeled with $i$.
}
\label{fig:meyer}
\end{figure}

The approximation of our model consists in retaining only those terms
where the $h_{ij}$ function appears only once
\begin{eqnarray}
\nonumber
\xi[ O_\eta(\bqu):\pon,\ptw,\hon,\htw]
& \longrightarrow &
\xi^1[ O_\eta(\bqu):\pon,\ptw,\hon,\htw] \\ & = & 
\langle \Phi; \pon,\ptw,\hon,\htw | \,O_{\eta}(\bqu)\, 
(1+ \sum_{i<j}\, h_{ij} ) \,|\Phi(0,0)  \rangle_L 
\label{eq:ximod1}
\\
\nonumber
& = & 
\langle \Phi; \pon,\ptw,\hon,\htw | 
\,O_{\eta}(\bqu)\, 
\sum_{1<j} h(r_{1,j}) \,|\Phi(0,0)  \rangle_L \, \\
& &+ \,
\langle \Phi; \pon,\ptw,\hon,\htw | \,O_{\eta}(\bqu)\, 
\sum_{1<i<j}\, h(r_{i,j}) \,|\Phi(0,0) \rangle_L 
\, .
\label{eq:ximod2}
\end{eqnarray}
This result has been obtained using a procedure analogous to that
adopted in Ref. \cite{co95} for the evaluation of the density
distribution, therefore the truncation of the expansion is done only
after the elimination of the unlinked diagrams. The Meyer-like
diagrams
considered in our work are presented in Fig. \ref{fig:meyer}. It is
evident that, in this case, the uncorrelated term, produced by the
$1$ in Eq. (\ref{eq:ximod1}), does not contribute. All the terms
necessary to calculate the cross section have been specified. The
detailed calculation of the various matrix elements for the one-body
current is presented in the Appendix A.

The $\Delta$ currents allow for the two-nucleon emission already at
the independent particle model level. We consider only this
contribution, therefore SRC and $\Delta$ currents are not directly
coupled in our model. They influence each other only via the
interference terms. 
\section{SPECIFIC APPLICATIONS}
\label{sec:app}
The interest in studying two-nucleon emission cross sections, is
related to the possibility of obtaining information on the SRC.
In order to identify correlation effects it is necessary
to keep under control all the other variables which can affect the
cross section.  For this reason, before studying the influence of the
SRC we shall discuss the sensitivity of the cross section to the
various theoretical inputs of our model.  The results of this
investigation will be presented only for the reaction
$^{16}$O(e,e'2p)$^{14}$C, but we have done calculations also for
$^{12}$C and $^{40}$Ca nuclei. The two-hole composition of the various
final states we have considered is given in Table \ref{tab:states}.

\begin{table}[hb]
\begin{center}
\begin{tabular}{|c|ccc|}
\hline
         &   $^{12}$C     &  $^{16}$O      & $^{40}$Ca \\
\hline
 $0_1^+$ & (1p3/2)$^{-2}$ & (1p1/2)$^{-2}$ & (1d3/2)$^{-2}$ \\
 $0_2^+$ &                & (1p3/2)$^{-2}$ & (2s1/2)$^{-2}$ \\
 $1^+$   &                & (1p1/2)$^{-1}$  (1p3/2)$^{-1}$ &  
                                           (1d3/2)$^{-1}$ (2s1/2)$^{-1}$\\
 $2_1^+$ & (1p3/2)$^{-2}$ & (1p1/2)$^{-1}$  (1p3/2)$^{-1}$ &
                                           (1d3/2)$^{-2}$ \\
 $2_2^+$ &                & (1p3/2)$^{-2}$ &  (1d3/2)$^{-1}$
                                              (2s1/2)$^{-1}$  \\ 
\hline
\end{tabular}
\end{center}
\caption{\small Two-hole compositions of the final states considered
         in our study.
}
\label{tab:states}
\end{table}

\subsection{Kinematics and No-Recoil Approximation}
\label{sec:nra}
The definition of the kinematics of the reaction under investigation
is far from being trivial. Energy and momentum conservation imply
\begin{eqnarray}
\label{eq:econ}
\omega &=& \epsilon_1+\epsilon_2 
         - \epsilon_{\hon}-\epsilon_{\htw} 
         + \frac {\bpi^2_r} {2 M_{A-2}} \, ,\\ 
\label{eq:qcon}
\bqu &=& \bpi_1 + \bpi_2 + \bpi_r
\, ,
\end{eqnarray}
where we have considered non relativistic kinematics and we have
indicated with $\bpi_r$ the recoil momentum of the $A-2$
residual nucleus. The other symbols have been already defined in the
previous sections. 

In our calculations we fixed $\omega$, $\bqu$, and $\bpi_2$.  By
selecting the $A-2$ nucleus final state also $\epsilon_{\hon}$ and
$\epsilon_{\htw}$ are fixed.  The cross sections are presented as a
function of $\gamma_{\pon}$. By solving the above system of equations
we obtain the values of $|\bpi_r|$ and $|\bpi_1|$ for each
$\gamma_{\pon}$.  This means that in the calculation of the nuclear
transition matrix elements, the radial integrals shown in Appendix
A, should be calculated for each value of $\gamma_{\pon}$.  From the
computational point of view, the situation improves enormously by
assuming $\bpi_r=0$, since in this case $|\bpi_1|$ becomes independent
from $\gamma_{\pon}$.  We call No-Recoil Approximation (NRA) this
approximation, consisting in setting to zero the kinetic energy and the
momentum of the residual $A-2$ nucleus.

\vspace*{1cm}

\begin{figure}[ht]
\hspace*{2.5cm}
\includegraphics[bb=50 200 700 700,angle=0,scale=0.7]
       {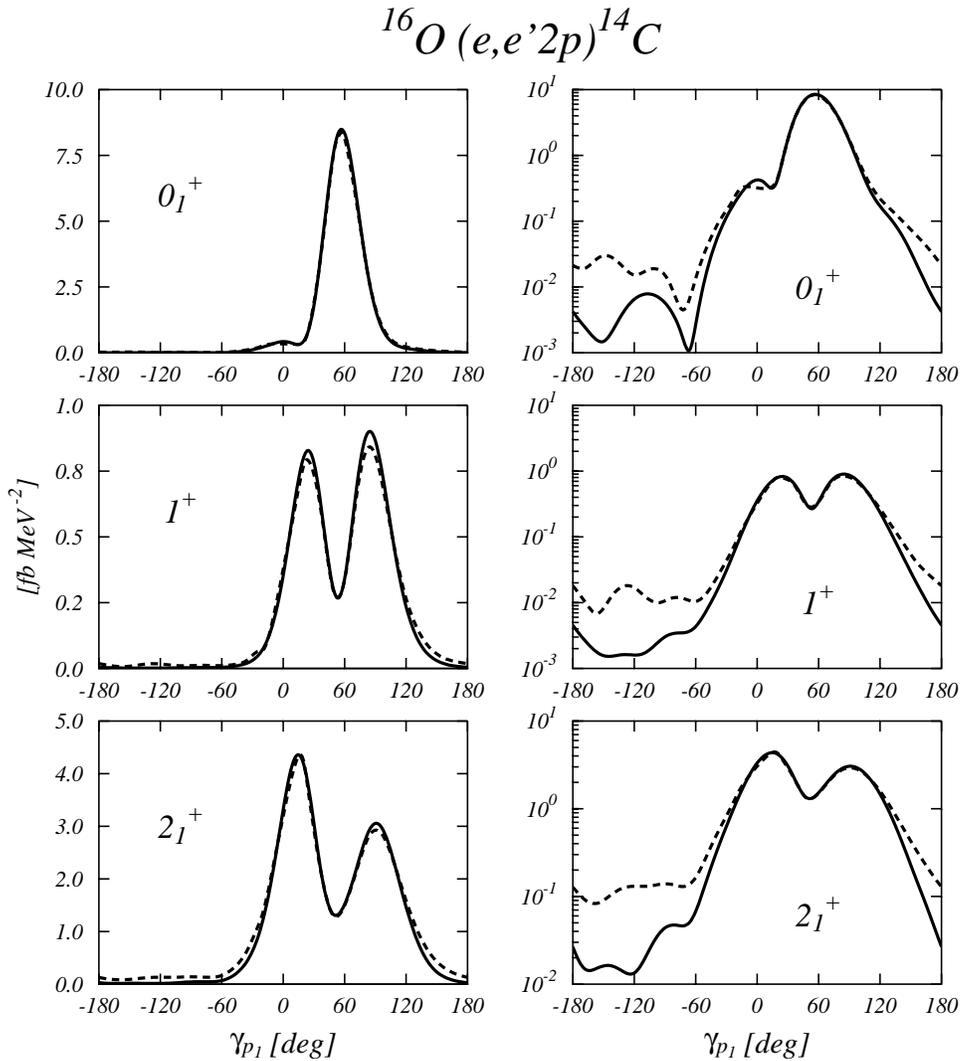}
\vspace*{0.8cm}
\caption{\small  Cross sections for the 
          $^{16}$O(e,e'2p)$^{14}$C reaction in the {\sl standard
          kinematics} (see Sect. \ref{sec:mf}). 
          The two-hole composition of the
          final states is given in Table \protect\ref{tab:states}.
          The dashed lines have been obtained by properly considering the
          recoil energy of the $^{14}$C nucleus. The full lines have
          been obtained within the NRA.
          The left and right panels show the same results in linear
          and logarithmic scale.
}
\label{fig:nra}
\end{figure}

In Fig. \ref{fig:nra} we compare the results of the NRA (full lines)
with those obtained by correctly treating the $A-2$ nucleus recoil
(dashed lines).  The reaction is $^{16}$O(e,e'2p)$^{14}$C and the
calculations have been done in what we shall define in Sect.
\ref{sec:mf} as {\sl standard kinematics}.  The left and right panels
show the same results in linear and logarithmic scale. This
presentation is done to show that the two results are very similar in
the peak region.  The differences are relevant only off the maximum
region, but they can be appreciated only in logarithmic scale since in
that region the cross section values fall by several order of
magnitude with respect to the peak values.

We have verified that, in the kinematics used in the calculations, the
recoil factor in Eq. (\ref{eq:rec}) differs from 1 at most by a 4\%.
Therefore the source of the difference is related to the change of the
single particle wave functions because of the change of
$\epsilon_1=\bpi_1^2/2m$.  In the kinematics under investigation this
energy can vary from 34.0 down to 22.6 MeV. It is interesting to
notice that, in the region where the cross section has its maximum, or
maxima, the results of the two calculations are rather similar. This
feature can be more easily understood by considering the emitted
nucleon wave functions as plane waves.  In this case by using
Eq. (\ref{eq:opft}) one can observe that the calculation of the matrix
element (\ref{eq:xi3}) is related to the Fourier transform of the
two-hole relative wave functions with respect to the variable $\bpi_r
= \bqu - (\bpi_1 + \bpi_2)$.  As it has been pointed out in
Refs. \cite{giu97,ryc96,ryc97}, the maximum value of this transform
occurs when the argument has its minimum value, possibly zero. For
this reason the results of the two calculations shown in
Fig. \ref{fig:nra} are rather similar in the region of the maxima.

The good performances of the NRA, at least in the region where the
cross section shows its maxima, gave us confidence about its use in our
calculations. All the results presented in the following sections have
been obtained in NRA.

\subsection{The mean field}
\label{sec:mf}
In this section we present the results of our study on the influence
of the mean field used to generate the single particle wave functions
on the (e,e'2p) cross sections. The investigation has been conducted
on $^{16}$O by fixing the correlation function and
changing the mean field.

As in the case of our previous works on (e,e'p) and ($\gamma$,p)
reactions \cite{mok01,ang02}, we used a real potential to generate the
hole wave functions, and an optical potential to generate the particle
wave functions.  With this choice the single particle wave functions
are not any more orthonormal. In our previous investigations
\cite{mok01,ang02} we have studied these non-orthogonality effects,
and we reached the same conclusions obtained in Ref. \cite{bof82}
where the problem has been first studied: for the case under
investigation and the kinematics we intend to explore, these effects
are negligible.

The parameters of the Woods-Saxon well used to generate the hole wave
functions are those given in Ref. \cite{ari96}. Associated to these
single particle wave functions there are SRC obtained by minimizing
the nuclear hamiltonian with the semi-realistic $S3$ interaction of
Afnan and Tang \cite{afn68}.  For our test cases we used the gaussian
form of the correlation function taken from \cite{ari96}:
\beq
f(r)=1-a\, e^{-b r^2}
\, ,
\label{eq:gaucor}
\eeq
with $a$=0.7 and $b$=2.2 fm$^{-2}$.  The particle wave functions have
been obtained by using the optical potential of Schwandt {\it et al.}
\cite{sch82}.  If not explicitly stated otherwise, these are the
parameters used in our calculations.

Since the number of variables is quite large we have chosen specific
kinematics to perform our test calculations. These kinematics, we call
them {\sl standard kinematics}, consist in fixing the electron
incoming energy $\epsilon$ = 800 MeV and the excitation energy
$\omega=100$ MeV. Then we fixed $\epsilon_2$ = 40 MeV and the
emission angle $\gamma_{\ptw}$ = 60$^{\rm o}$. The energy of the other
proton is obtained from Eq. (\ref{eq:econ}) by setting $\bpi_r=0$,
that is in the NRA, and the cross section are presented as a function
of $\gamma_{\pon}$, the emission angle. We always work
in coplanar kinematics, i.e. $\phi_{\pon}$ and $\phi_{\ptw}$ are zero.
We recall that we have taken the angle $\gamma_{\pon}$ to be positive on
the opposite side of the scattered electron with respect to $\bqu$,
while $\gamma_{\ptw}$ is taken positive on the same side of the scattered
electron.  We used this {\sl standard kinematics} in $^{16}$O as a
reference calculation.

\begin{figure}[ht]
\hspace*{2.5cm}
\vspace*{1cm}
\includegraphics[bb=50 200 500 700,angle=0,scale=0.7]
       {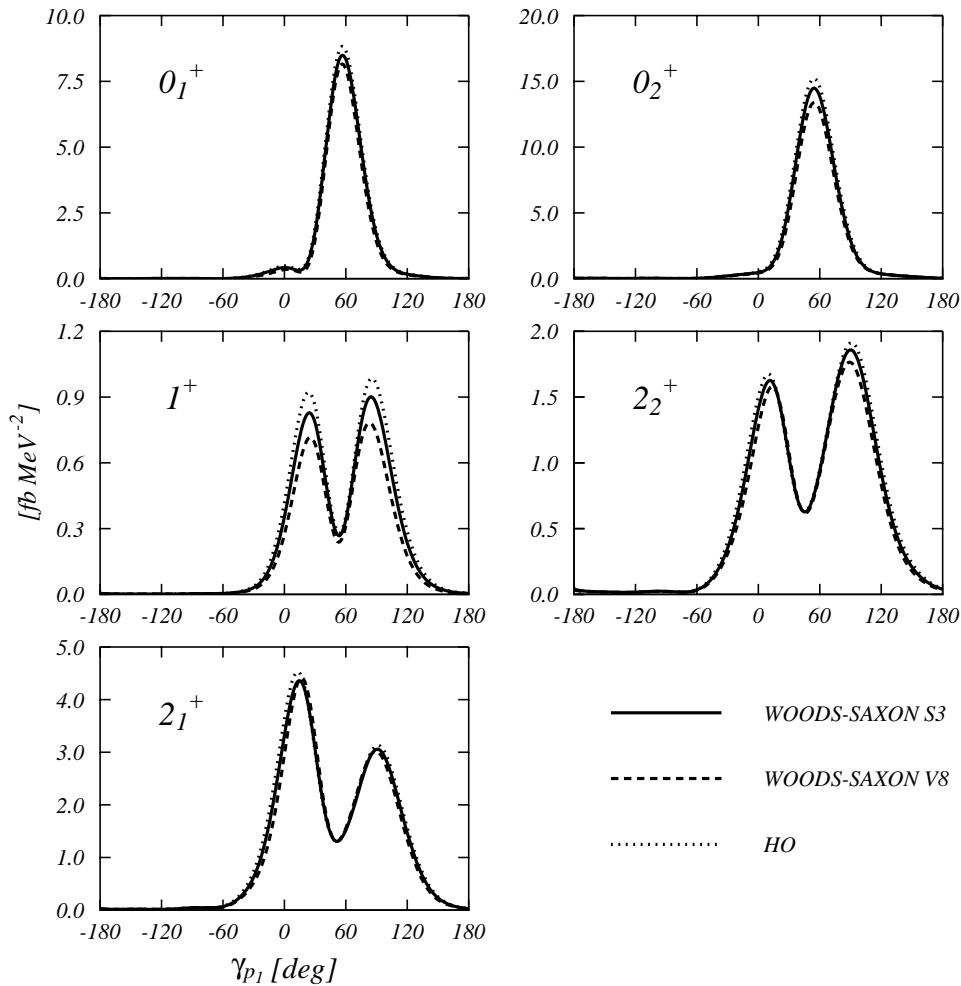}
\vskip .0 cm 
\caption{\small  Cross sections calculated in {\sl standard
       kinematics} with different mean field potentials generating
       the hole wave functions. The full lines show the standard
       results obtained by using the mean-field potential of
       Ref. \protect\cite{ari96}. The dashed lines have been obtained
       with the mean field potential of Ref. \protect\cite{fab00}. The
       dotted lines are results obtained with harmonic oscillator wave
       functions. 
}
\label{fig:ho}
\end{figure}

A first test has been done on the sensitivity of the results to the
changes of the hole wave functions.  An example of the results we have
obtained is shown in Fig. \ref{fig:ho}. The standard results are shown
by the full lines. The dashed lines have been obtained with the hole
wave functions generated by the Woods-Saxon potential whose parameters
are given in Table V of Ref. \cite{fab00}. These parameters have been
used in a Fermi hypernetted chain calculation done with the Argonne
$V8'$ nucleon-nucleon interaction. In this case the spin orbit term of
the mean-field potential has been set to zero.  The dotted lines have
been obtained by using harmonic oscillator single particle wave
functions.  The sensitivity of various results to these changes is
limited: we found a maximum variation of 4\%, except for the case of
the 1$^+$ state. This state is, however, out of the systematics, since
it is dominated by the $\Delta$ currents, as we shall discuss in more
detail in Sect. \ref{sec:delta}.

We should remark that all the mean-field potentials used in these test
calculations reproduce the root mean square charge radii. Furthermore,
the single particle energies have been kept constant in all the
calculations.  In our calculations, the hole single particle energies
are used to determine the particle energies, see Eq.
(\ref{eq:econ}).  The particle wave functions are calculated by
solving the single particle Schr\"odinger equation with an energy
dependent optical potential for a continuum wave. Changes in the
energy of the particle could strongly modify the single particle
hamiltonian, then the particle wave functions.

\begin{table}[ht]
\begin{center}
\begin{tabular}{|c|rrrr|}
\hline
                & 1d5/2 & 2s1/2 & 1p1/2 & 1p3/2 \\
\hline
 $^{12}$C       &       &  &         & -18.0  \\
 $^{16}$O       &       &  & -12.7  & -16.9   \\
 $^{40}$Ca      & -8.7 & -19.3   &  &   \\
\hline
\end{tabular}
\end{center}
\caption{\small Single particle energies, in MeV, for the states of
                interest in our calculations. These energies have been
                obtained using the Woods-Saxon parameterization given
                in \protect\cite{ari96}.
 }
\label{tab:spe}
\end{table}

In our {\sl standard kinematics} calculations we used the hole
energies obtained by solving the single particle Schr\"odinger
equation with the ground state Woods-Saxon potential.  In the cases of
our interest these single particle energies are given in Table
\ref{tab:spe}.  In this approach we do not consider the fact that a
residual nucleon-nucleon interaction rearranges the mean field in the
$A-2$ nucleon system such that its energy is not any more the energy of
the A nucleon system minus the two single particle energies. A way to
take into account this fact is to use the experimental binding
energies of the $A-2$ nucleus, the $^{14}$C in the specific case, and
the energies of the first excited states which should be interpreted
in terms of single particle excitations. This is the approach used by
Giusti and Pacati \cite{giu97}.

\vspace*{1cm}

\begin{figure}[hb]
\hspace*{2.5cm}
\includegraphics[bb=50 200 500 700,angle=0,scale=0.7]
       {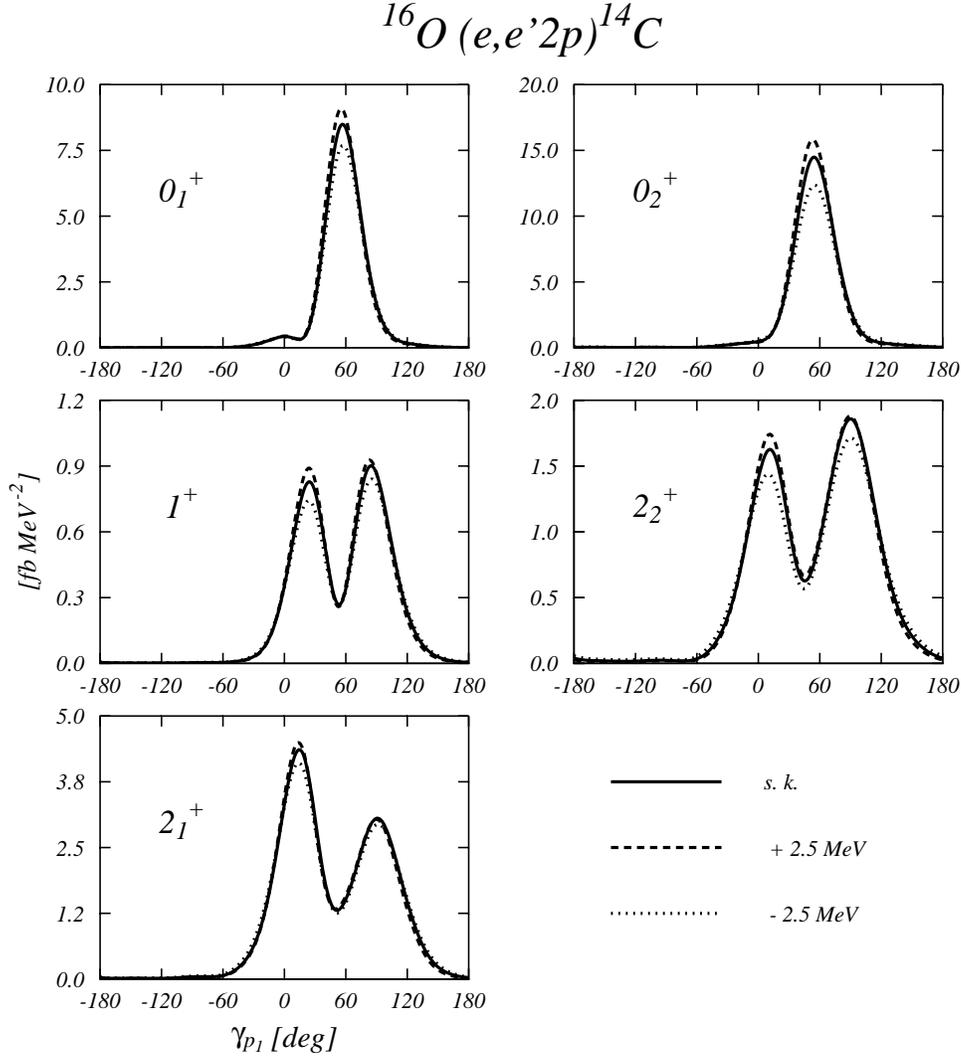}
\vskip 1 cm 
\caption{\small Comparison between cross sections calculated in {\sl
       standard kinematics} (full lines) and those obtained by
     increasing (dashed lines)  and reducing (dotted lines)
     $\epsilon_1$ by 2.5 MeV. 
}
\label{fig:spe}
\end{figure}

From the pragmatic point of view the two methods of determining the
hole single particle energies introduce an uncertainty of about 2.5
MeV. This uncertainty is reflected on the energies of the emitted
particle and on their wave functions. The final uncertainties are shown
in Fig. \ref{fig:spe} where the results of the {\sl standard
calculations} are compared with those obtained by increasing and
lowering $\epsilon_1$ by 2.5 MeV.

\begin{figure}[ht]
\hspace*{2.5cm}
\includegraphics[bb=50 200 500 700,angle=0,scale=0.7]
       {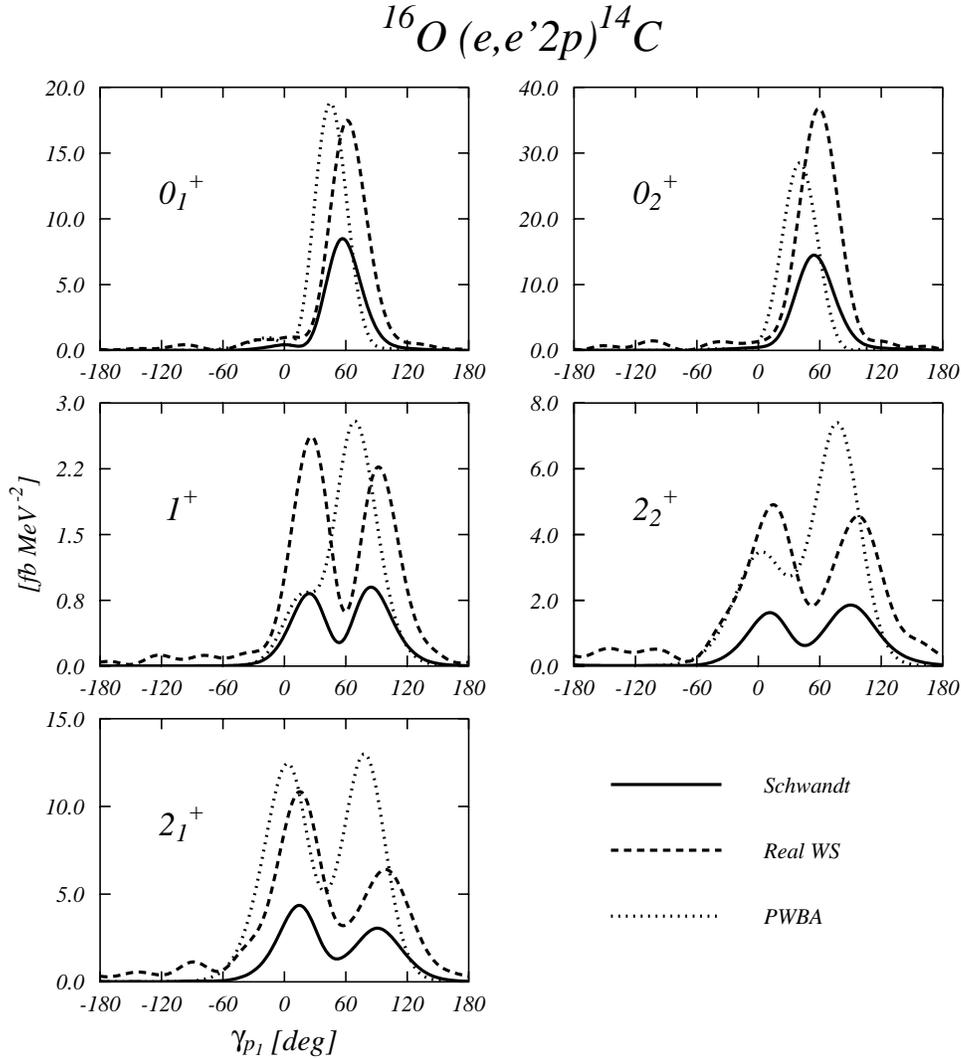}
\vskip 1.0 cm 
\caption{\small Cross sections calculated in {\sl standard
       kinematics}.  The standard results
       (full lines) are compared with those obtained by using
       plane waves (dotted lines) and real Woods-Saxon single particle wave
       functions (dashed lines).    
}
\label{fig:pwba}
\end{figure}

The larger variation of the cross section obtained in Fig.
\ref{fig:spe} with respect to those of Fig. \ref{fig:ho} indicates the
great sensitivity of the results to the particle wave functions. In
Fig. \ref{fig:pwba} we compare the results of the {\sl standard
calculation} with those obtained by using plane waves for the particle
wave functions (dotted lines) and with those obtained by using the
same real Woods-Saxon potential utilized to describe the hole states
(dashed lines). It is evident the large difference between the various
results and it appears clear the effect of the complex optical
potential whose imaginary part quenches the cross section.

\vspace*{1cm}

\begin{figure}[ht]
\hspace*{2.5cm}
\includegraphics[bb=50 200 500 700,angle=0,scale=0.7]
       {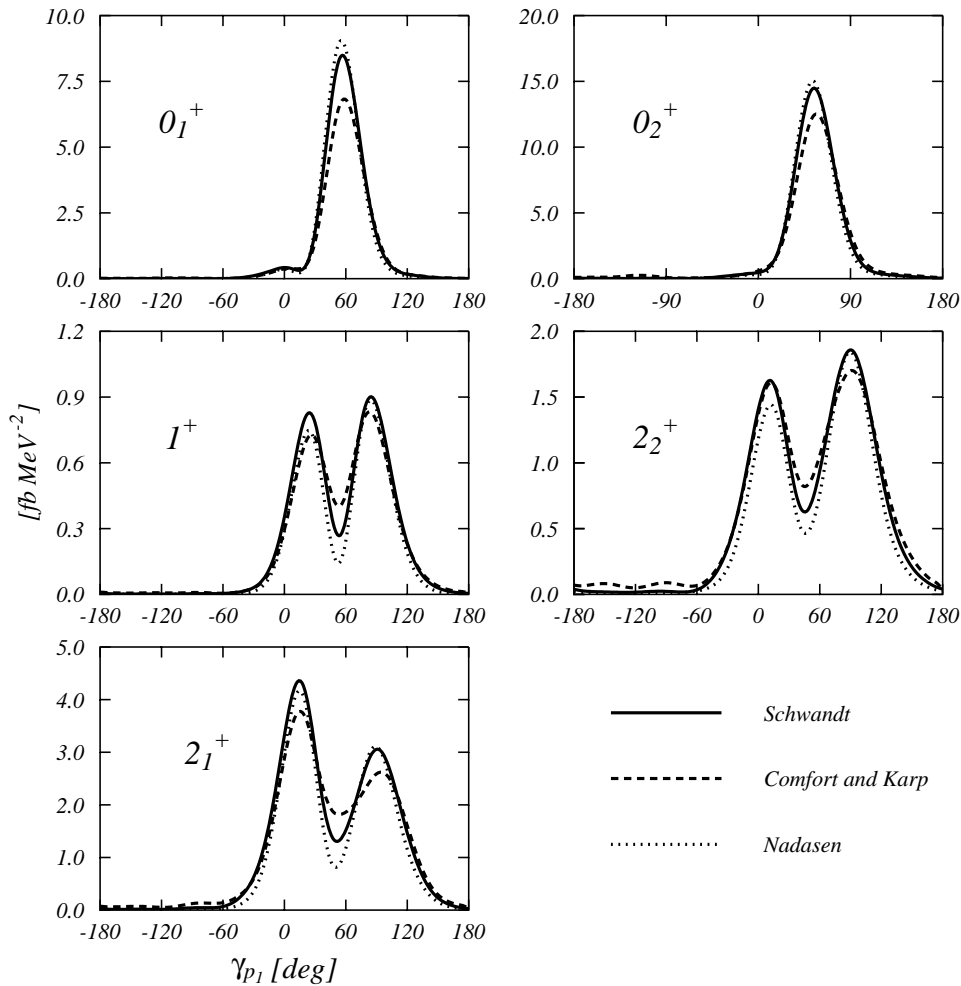}
\vskip 1.0 cm 
\caption{\small Cross sections calculated in {\sl standard
       kinematics}. The standard results
       (full lines) are compared with those obtained with the optical
       potential of Comfort and Karp \protect\cite{com80}
       (dotted lines) and that of Nadasen {\it et al.} \protect\cite{nad81}
       (dashed lines).
}
\label{fig:opt}
\end{figure}

The sensitivity of our calculations to the choice of the optical
potential is shown in Fig. \ref{fig:opt} where the cross sections have
been calculated by using the optical potentials of Schwandt {\it et al.} 
\cite{sch82} (the {\sl standard calculations}), Nadasen {\it et
al.} \cite{nad81} and Comfort and  Karp \cite{com80}. Though the
modifications of the cross section are less pronounced than in
Fig. \ref{fig:pwba}, they can become as large as 18\%.

We can conclude this section by saying that the various 
plausible choices of the mean-field parameters can produce
uncertainties of about 20\% - 30\% on the maxima of the cross
sections. 

\subsection{The two-body current}
\label{sec:delta}
We have already pointed out that the presence of SRC is not the only
mechanism inducing the emission of two protons. Meson exchange
currents of the type shown in Fig. \ref{fig:delta} can also contribute
to this process. For the purpose of studying SRC the presence of these
two-body currents is disturbing, therefore, it is important to find
kinematics where the $\Delta$ currents contributions are negligible 
with respect to the SRC effects.

In this section we discuss the sensitivity of our results to the
presence of the $\Delta$ currents.  There are various aspects of the
description of electromagnetically induced $\Delta$ excitation that
are still largely debated \cite{chr90,gil97}.  It is not our aim to
enter in this discussion. We have considered a $\Delta$ current model
commonly used in the literature, we have applied it to two-nucleon
emission processes, and we have analyzed the sensitivity of our
results to the changes of the parameters of the model and of the
kinematics.

\begin{figure}
\hspace*{2.5cm}
\includegraphics[bb=50 200 500 700,angle=0,scale=0.7]
       {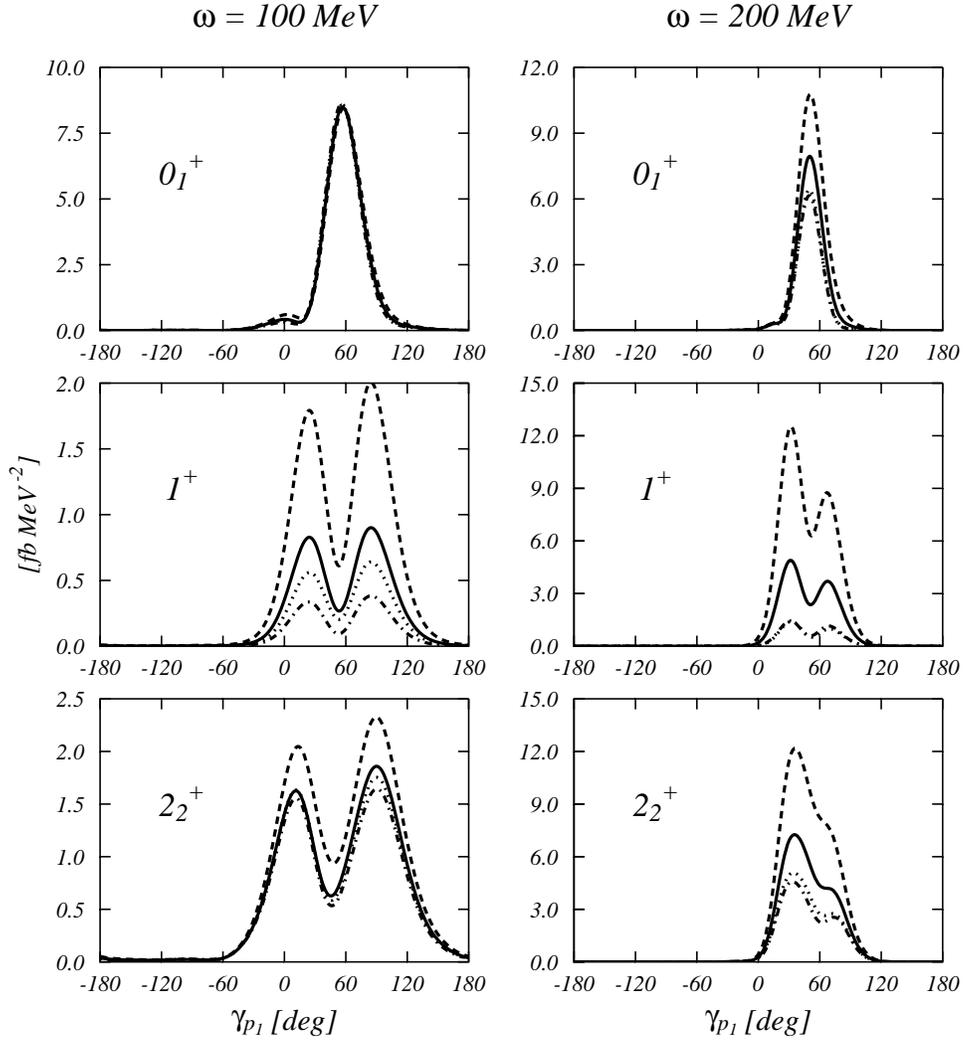}
\vskip 1.0 cm 
\caption{\small Cross sections for the $^{16}$O(e,e'2p)$^{14}$C
       process  calculated 
       using different values of the $\Delta$ current coupling
       constants (see Table \protect\ref{tab:delta}). The full lines
       have been obtained with the parameters of
       Ref. \protect\cite{ama98}, and these are the values used in our
       {\sl standard} calculations. The dashed and dashed dotted lines
       have been obtained with the parameters of
       Refs. \protect\cite{giu97} and \protect\cite{ryc97} respectively.
       Dotted curves correspond to the {\sl standard} calculations but
       using the static propagator for the $\Delta$ currents. 
}
\label{fig:d100}
\end{figure}

We present in Fig. \ref{fig:d100} some specific result obtained in
{\sl standard kinematics} for $\omega=100$ MeV and $\omega=200$ MeV
with the aim of discussing two different features regarding the
$\Delta$ currents. The first one is related to the sensitivity of the
results to the various coupling constants defined in Eq.
(\ref{eq:delta}). The second one concerns the relative importance of
SRC and $\Delta$ currents with respect to the change of the excitation
energy.

Out of the three constants $f_{\pi N N}$, $f_{\pi N \Delta }$ and
$f_{\gamma N \Delta}$, only the first one is rather well defined by
the experimental data. Its commonly accepted value is $f^2_{\pi N N}/4
\pi$=0.079. The values of the other two constants are not so precisely
known.  In our {\sl standard calculations} we used the values taken
from \cite{ama94}. In Figs. \ref{fig:d100} we have also used the
values adopted in Refs.  \cite{giu97} and \cite{ryc97} (see Table
\ref{tab:delta}).

\begin{table}
\begin{center}
\begin{tabular}{|r|rrr|}
\hline
                            & AMA & GIU & RYC  \\
\hline
 $f_{\gamma N \Delta}$      & 0.299 & 0.373 & 0.120  \\
 $f_{\pi N \Delta}$         & 1.69  & 2.15  & 2.15  \\
\hline
\end{tabular}
\end{center}
\caption{\small Values of the parameters used in
                            Eq. (\protect\ref{eq:delta}). 
        The AMA, GIU and RYC values are from Refs. \protect\cite{ama94},
          \protect\cite{giu97} and   \protect\cite{ryc97} respectively.
}
\label{tab:delta}
\end{table}

We have also tested the validity of the static approximation in the
description of the $\Delta$ propagator. This approximation consists
essentially in setting $G_A=G_B=1/(m_\Delta-m_n)$ in
Eq. (\ref{eq:gi}). This is the approximation we have used in our
previous works \cite{co01,ang02,ang01}. The dotted lines of Fig.
\ref{fig:d100} show the results obtained with the static
approximation.

\begin{figure}[ht]
\hspace*{2.5cm}
\includegraphics[bb=50 200 500 700,angle=0,scale=0.7]
       {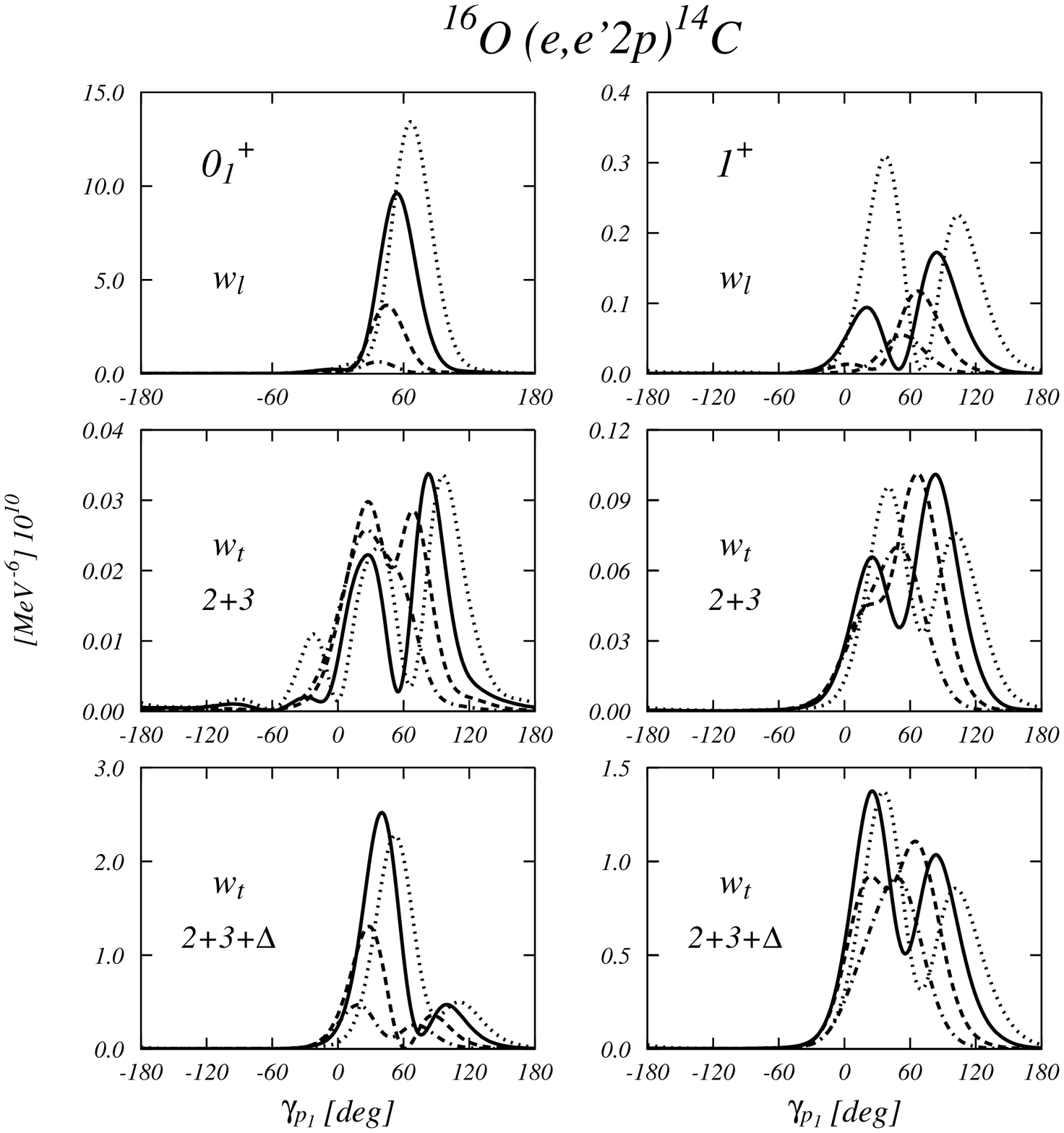}
\vskip 1.0 cm 
\caption{\small Longitudinal ($w_l$) and transverse ($w_t$) responses, 
       Eqs. (\protect\ref{eq:wl}) and (\protect\ref{eq:wt}), calculated in
       standard kinematics, for various values of the momentum
       transfer: dotted lines 300 MeV/c, full lines 400 MeV/c, dashed lines
       500 MeV/c and dashed-dotted lines 600 MeV/c. The two central
       panels have been calculated without $\Delta$ currents
       contribution, while the curves of the lower panels include
       also this contribution.
}
\label{fig:resp}
\end{figure}

We observe that at $\omega=100$ MeV the $0_1^+$ state shows scarce
sensitivity to the changes of the values of the $\Delta$ currents
coupling constants. More sensitive is the $2_2^+$ state, but certainly
the most sensitive one is the $1^+$. These different degrees of
sensitivity are connected to the relative importance of the $\Delta$
current contribution with respect to the SRC in the various excited
states.  As already pointed out in Refs.  \cite{giu97} and
\cite{ryc97}, the $1^+$ state is strongly dominated by the $\Delta$
currents, therefore any change in the coupling constants produces
large modifications of the cross section. The small sensitivity of the
$0_1^+$ state is due to the fact that its excitation is mainly due to
the longitudinal response (see Fig. \ref{fig:resp}), while the
$\Delta$ acts on the transverse response only. The $\Delta$
contribution is not negligible for the $2_1^+$ state and it becomes
dominant for the $1^+$.

The results of the static approximation have been obtained by using
the constants of Ref. \cite{ama94}. This approximation is valid when
its results are closed to those shown by the full lines. At 
$\omega$=100 MeV this is the case for all the states we have
investigated but the $1^+$. We do not show  the
results for the $0_2^+$ and $2_1^+$ states that present analogous 
features of $0_1^+$ and $2_2^+$ results.

The situation changes for the excitation energy $\omega$=200 MeV.  Now
we are in the resonant region of the $\Delta$ and here all the states
are strongly sensitive to the different values of the coupling
constants. The transverse response dominates also in the $0_1^+$
state. The results of the static approximations are quite far from the
full lines, clearly showing the inadequacy of this approximation.
It is interesting to notice that the static approximation produces
results rather similar to those obtained with the coupling constants
of Ref. \cite{ryc97}.

\begin{figure}[ht]
\hspace*{2.5cm}
\includegraphics[bb=50 200 500 700,angle=0,scale=0.7]
       {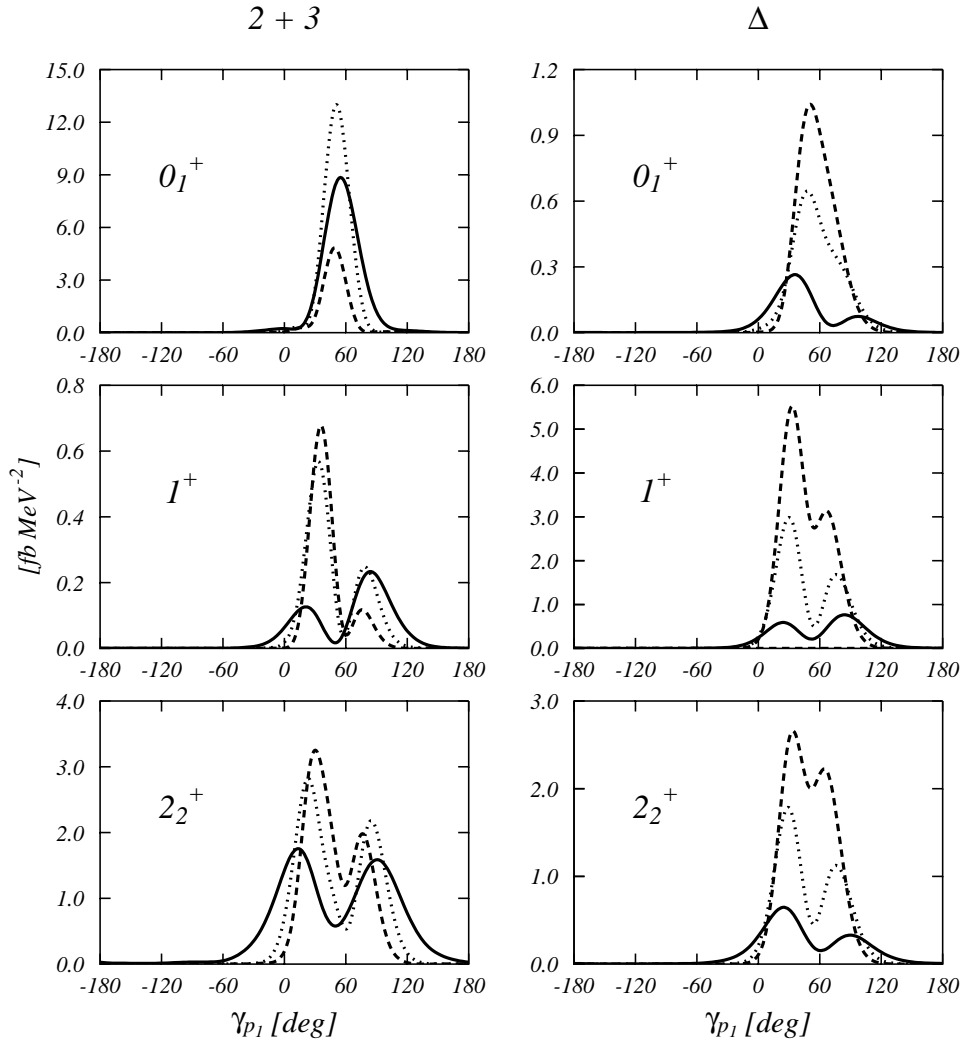}
\vskip 1.0 cm 
\caption{\small  Cross sections for the $^{16}$O(e,e'2p)$^{14}$C
       process  calculated 
       for three values of the excitation energy. Full lines
       $\omega$=100 MeV, dotted lines $\omega$=150 MeV, dashed lined 
       $\omega$=200 MeV. The left panels show the results obtained by
       considering only the one-body currents plus the SRC. 
       The right panels show
       the cross sections obtained with $\Delta$ currents only.
}
\label{fig:edep}
\end{figure}

We have investigated the energy and the momentum dependence induced by
the SRC and by the $\Delta$ currents. The results of our study about
the energy dependence are summarized in Fig. \ref{fig:edep}, where in
the left panels we show the cross sections calculated using only the
one-body currents plus SRC, while in the right panels we show the
results obtained with the $\Delta$ currents alone.  The comparison
between the left and right panels for the $1^+$ state, clearly show
that the calculations with the $\Delta$ currents produce cross
sections much larger than those obtained with one-body currents only.
This confirms that the excitation of this state is dominated by the
$\Delta$ currents dynamics.  The same analysis requires a more careful
discussion in the case of the $0_1^+$ state. At $\omega =$ 100 and 150
MeV, the cross sections obtained with the one-body currents are larger
than those calculated with the $\Delta$ currents alone by a factor
20-30.  At $\omega =$ 200 MeV the difference between the two results
is about five times smaller.  This indicates that, even for the $0_1^+$
state, the contribution of the $\Delta$ currents becomes important
when the excitation energy is in the $\Delta$ resonance region.  For
the $2_2^+$ state the contributions of the one-body and of the
$\Delta$ currents have the same order of magnitude.

The SRC do not show specific energy dependence. We have verified that
by calculating all the $^{16}$O final states of Table
\ref{tab:states}. In the example shown in Fig. \ref{fig:edep} one can
observe that in the $0_1^+$ state the cross sections at 200 MeV is
smaller than at 150 MeV, while the size of the cross sections for
these excitation energies is almost the same for the $2_2^+$ state.
The situation is different for the case of the $\Delta$ currents whose
contribution grows with the energy since it enters in its resonance
region.

Some results of our investigation of the momentum transfer dependence
are shown in Fig. \ref{fig:resp}. In this figure, instead of the cross
sections, we present the longitudinal and transverse responses, Eqs.
(\ref{eq:wl}) and (\ref{eq:wt}), to get rid of the trivial $\bqu$
dependent terms in the kinematics factors.  We show the responses for
the $0_1^+$ and $1^+$ final states.

The $0_1^+$ state is dominated by the longitudinal response. The shape
of the angular distribution of $w_l$ is not modified by the change of
$|\bqu|$, however its peak value is reduced with increasing $|\bqu|$.
The values of $w_t$ without $\Delta$ currents are so small to be
negligible. The inclusion of the two-body currents enhances the
transverse response. It has been shown in Ref. \cite{giu98} that, in
this state, the emission of the two nucleons is dominated by a relative
$s$-wave. The $|\bqu|$ dependences of longitudinal and transverse
responses are quite different. At $|\bqu|$ = 300 and 400 MeV/c, the
longitudinal responses are much larger than the transverse ones. The
peak values of the responses become smaller with increasing $|\bqu|$,
but the decrease of $w_l$ is steeper than that of $w_t$. At $|\bqu|$=
600 MeV/c we are in a situation where the size of the two responses
is about the same.  In this kinematics the presence of the
$\Delta$ currents noticeably affect the full cross section.

The situation is different for the $1^+$ state. This state is
dominated by the $\Delta$ currents, therefore the transverse response
is the most important one. The emission of the two protons is ruled by
a relative $p$-wave \cite{giu98}, which produces an angular
distribution with two peaks. These two peaks are clearly visible when
the kinematics verifies $\bqu \approx \bpi_1+\bpi_2$.  If
$\bqu \gg \bpi_1+\bpi_2$ for any value of $\gamma_{\pon}$,
the two peaks merge as it is shown by the dashed and dashed dotted
lines. The change of $|\bqu|$ does not modify sensitively the size of
the peaks.

To summarize the results of this section, we may say that, in order to
minimize the effects of the $\Delta$ currents, one has to work at
excitation energies far from the resonance peak, at relatively low
values of the momentum transfer ($|\bqu| \leq$ 500 MeV/c), and
furthermore one should not consider the $1^+$ state.

\subsection{The Short Range Correlation}
\label{sec:src}
After having studied the sensitivity of our results to the theoretical
inputs and to the presence of the $\Delta$ currents, we arrive now to
the main point of our investigation: the sensitivity of the cross
section to the details of the correlation function. It is obvious that
the use of very different correlations would produce rather different
results, since $f(r)$ is the key ingredient of the calculation. The
point is to investigate the sensitivity of the cross section to
relatively small changes of realistic correlation functions.

\begin{figure}[ht]
\hspace*{1.cm}
\includegraphics[bb=50 320 360 750,angle=90,scale=0.9]
       {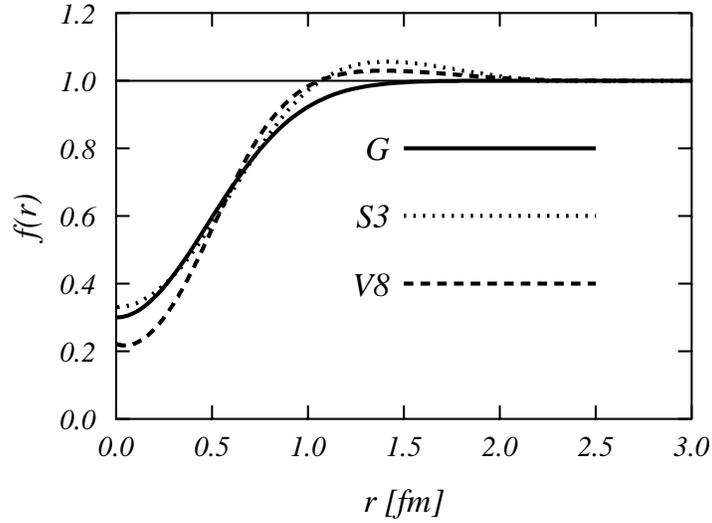}
\vskip -1.5 cm 
\caption{\small  Correlation functions used in our calculations. The
       line labeled $G$
       indicates the gaussian correlation function of
       Ref. \protect\cite{ari96} and that labeled $S3$ the Euler
       correlation 
       function of the same reference. With $V8$ we show the scalar
       term of the state dependent correlation function of
       Ref. \protect\cite{fab00}.        
}
\label{fig:corr}
\end{figure}

For our investigation we have considered the purely scalar correlation
functions $f(r)$ shown in Fig. \ref{fig:corr}.  The correlation
function $G$, we have used up to now, and the $S3$ correlation
function, are taken from Ref. \cite{ari96}. They have been fixed by
minimizing the energy functional calculated with a nuclear hamiltonian
containing the Afnan and Tang nucleon-nucleon interaction
\cite{afn68}.  The two minimizations have been done by considering
respectively a gaussian type correlation ($G$) and the Euler procedure
described in that reference ($S3$).  The correlation labeled as $V8$ in
the figure, is the scalar part of the state dependent correlation used
in Ref. \cite{fab00}, where the hamiltonian was built by using the
nucleon-nucleon Argonne $V8'$ interaction plus the Urbana IX three-body
force.  The three correlation functions differ for few details. The $S3$
and $V8$ overshoot the asymptotic value of 1 in the region between 1 and
2 fm. The $V8$ correlation function has a lower minimum than the other
two.

\begin{figure}[ht]
\vspace*{-1cm}
\hspace*{2.cm}
\includegraphics[bb=50 200 500 700,angle=90,scale=0.7]
       {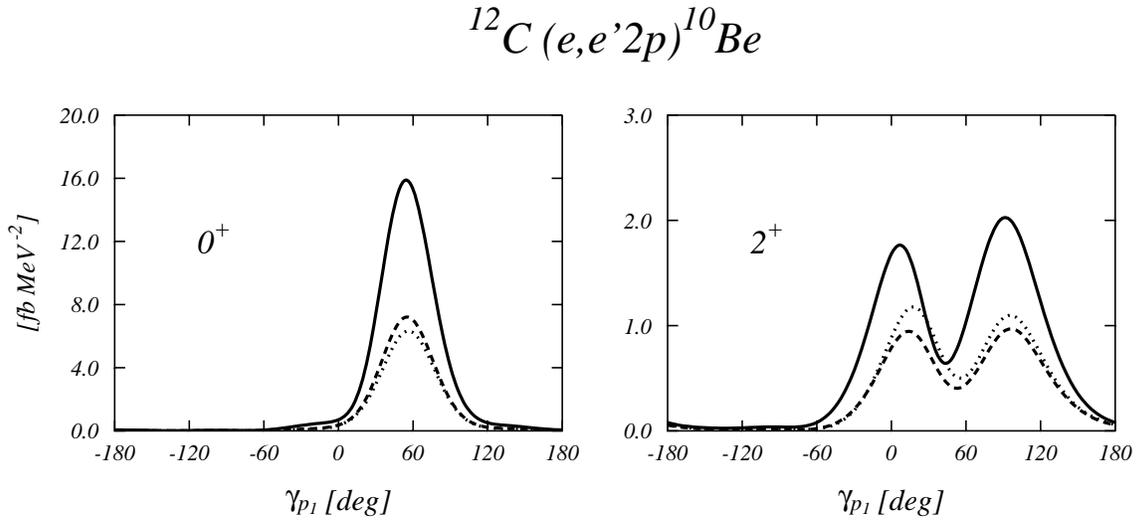}
\vskip -3.0 cm 
\caption{\small $^{12}$C(e,e'2p)$^{10}$B cross section  
  in the {\sl standard kinematics} for the emission of the two proton
  from the 1p3/2 level of the target nucleus. The full, dotted and
  dashed lines, have been obtained by using respectively the $G$, $S3$
  and $V8$ correlation functions of Fig. \protect\ref{fig:corr}.
  }
\label{fig:c12}
\end{figure}

\begin{figure}[ht]
\hspace*{2.5cm}
\includegraphics[bb=50 200 500 700,angle=0,scale=0.7]
       {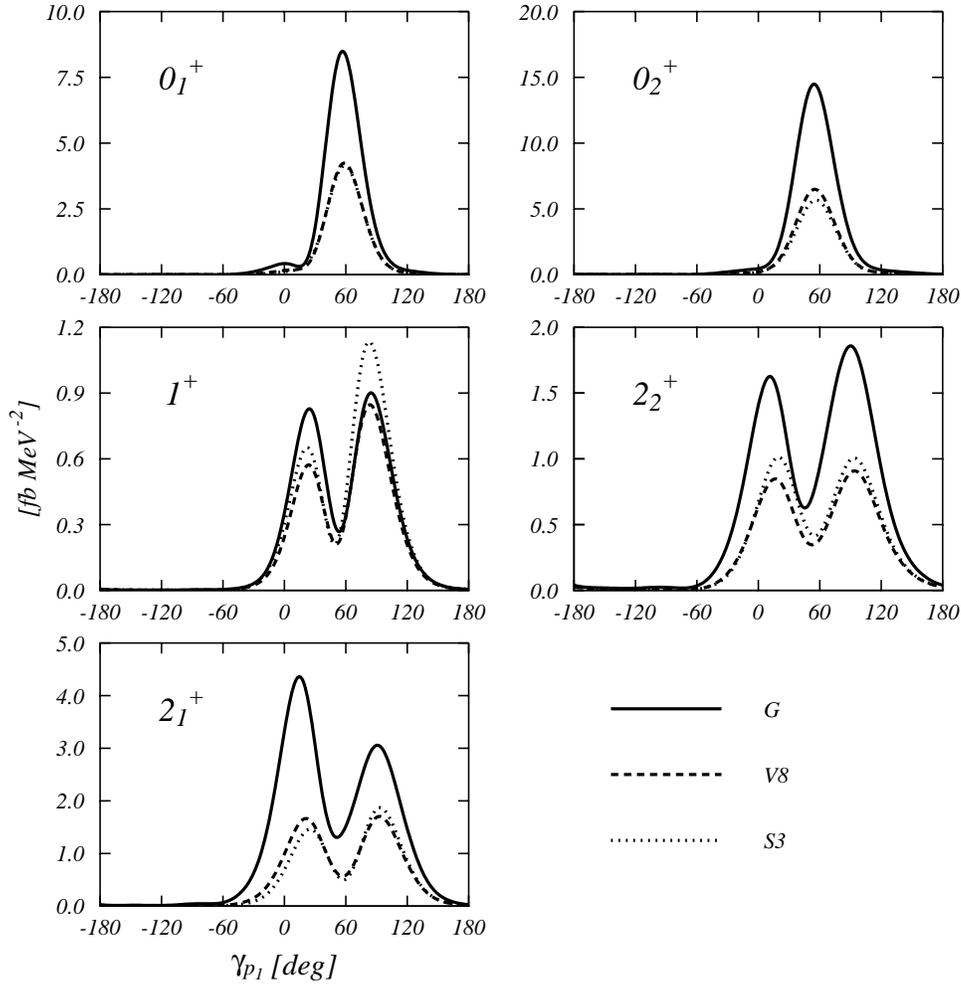}
\vskip 1.0 cm 
\caption{\small $^{16}$O(e,e'2p)$^{14}$C cross sections calculated in
       {\sl standard kinematics}
       with the three correlation functions shown in
       Fig. \protect\ref{fig:corr}. 
}
\label{fig:s3v8}
\end{figure}

\begin{figure}[ht]
\hspace*{2.5cm}
\includegraphics[bb=50 200 500 700,angle=0,scale=0.7]
       {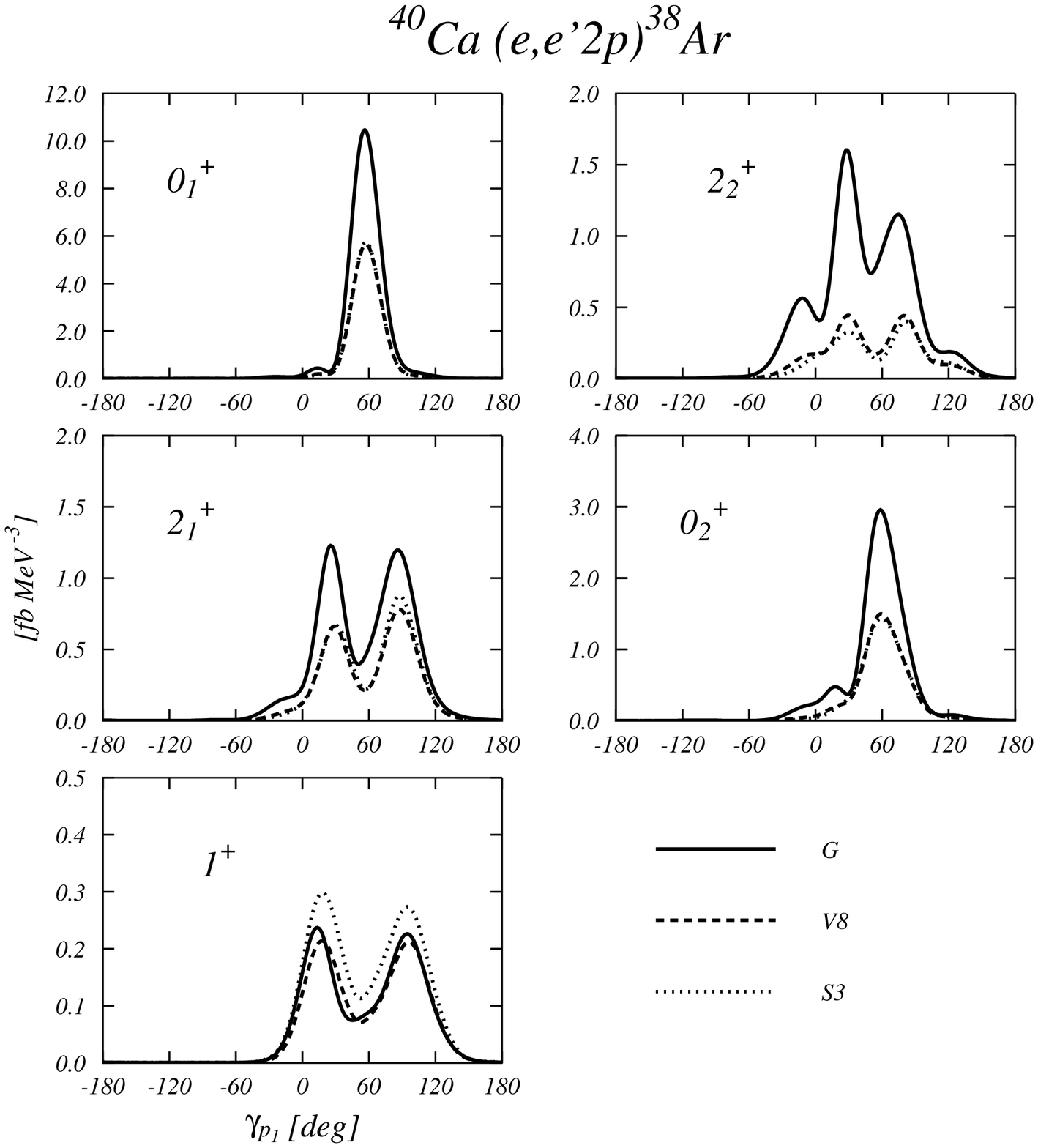}
\vskip 1.0 cm 
\caption{\small
  $^{40}$Ca(e,e'2p)$^{38}$Ar cross section calculated   
  in the {\sl standard kinematics}. The full, dotted and
  dashed lines, have been obtained by using respectively the $G$, $S3$
  and $V8$ correlation functions of Fig. \protect\ref{fig:corr}.}
\label{fig:ca40}
\end{figure}

In the Figs. \ref{fig:c12}, \ref{fig:s3v8} and \ref{fig:ca40} we show
the cross sections calculated in {\sl standard kinematics} with the
three correlation functions shown in Fig. \protect\ref{fig:corr} for
three target nuclei: $^{12}$C, $^{16}$O and $^{40}$Ca. The list of the
two-hole wave functions describing the final states of the $A-2$ rest
nuclei is given in Table \ref{tab:states}.

Apart from details related to the specific nucleus and the final state
considered, the general trend shown by these results is that the
shapes of the angular distributions are not modified by the
correlations. It is rather the size of the cross sections in their
maxima which has changed.  This indicates that the shapes of the
angular distributions are ruled by the kinematics and by the angular
momentum coupling between the hole states and the allowed partial
waves where the two particles can be emitted. This has been widely
discussed in \cite{giu98} where a decomposition of the partial wave in
terms of center of mass and relative motion of the emitted nucleon
pair has been done.

The cross sections obtained by using the $V8$ and $S3$ correlations
are rather similar and they are roughly a factor two smaller than
those obtained with the gaussian correlation, which we took as {\sl
standard} results. This effect is certainly larger than the
theoretical uncertainties estimated in section \ref{sec:mf}. 
The $1^+$ states are out of systematics because they are dominated by the
$\Delta$ currents.

\vspace*{1cm}

To investigate the sources of these differences we have done
calculations with rather schematic correlations which allowed us to
switch on and off various characteristics of the correlation
functions. To this purpose
we used step function correlations of the type
\beq
f(r) = \left\{ \begin{array} {ll}
         A \, ,& 0 \leq r < d_1 \\
         B \, ,& d_1 \leq r <d_2 \\
         1.0 \, , &  d_2 \leq r 
                \end{array}
       \right.
\label{eq:scat}
\eeq
We have changed the values of the $A$,$B$,$d_1$ and $d_2$ parameters
to produce four different correlations. The values of the parameters
used in our calculations are given in Table \ref{tab:scat}.

\vspace{1cm}

\begin{table}[ht]
\begin{center}
\begin{tabular}{|c|cccc|}
\hline
                   &  $A$  & $B$ & $d_1$ [fm] & $d_2$ [fm]  \\
\hline
 CSC1              &  0.2  & 1.0 & 1.0 & 1.0   \\
 CSC2              &  0.1  & 1.0 & 1.0 & 1.0   \\
 CSC3              &  0.2  & 1.1 & 1.0 & 1.3   \\
 CSC4              &  0.2  & 1.0 & 0.9 & 0.9   \\
\hline
\end{tabular}
\end{center}
\caption{\small Parameters of the schematic correlation functions 
as given into Eq. \ref{eq:scat}. 
}
\label{tab:scat}
\end{table}

\begin{figure}[ht]
\hspace*{2.5cm}
\includegraphics[bb=50 200 500 700,angle=0,scale=0.7]
       {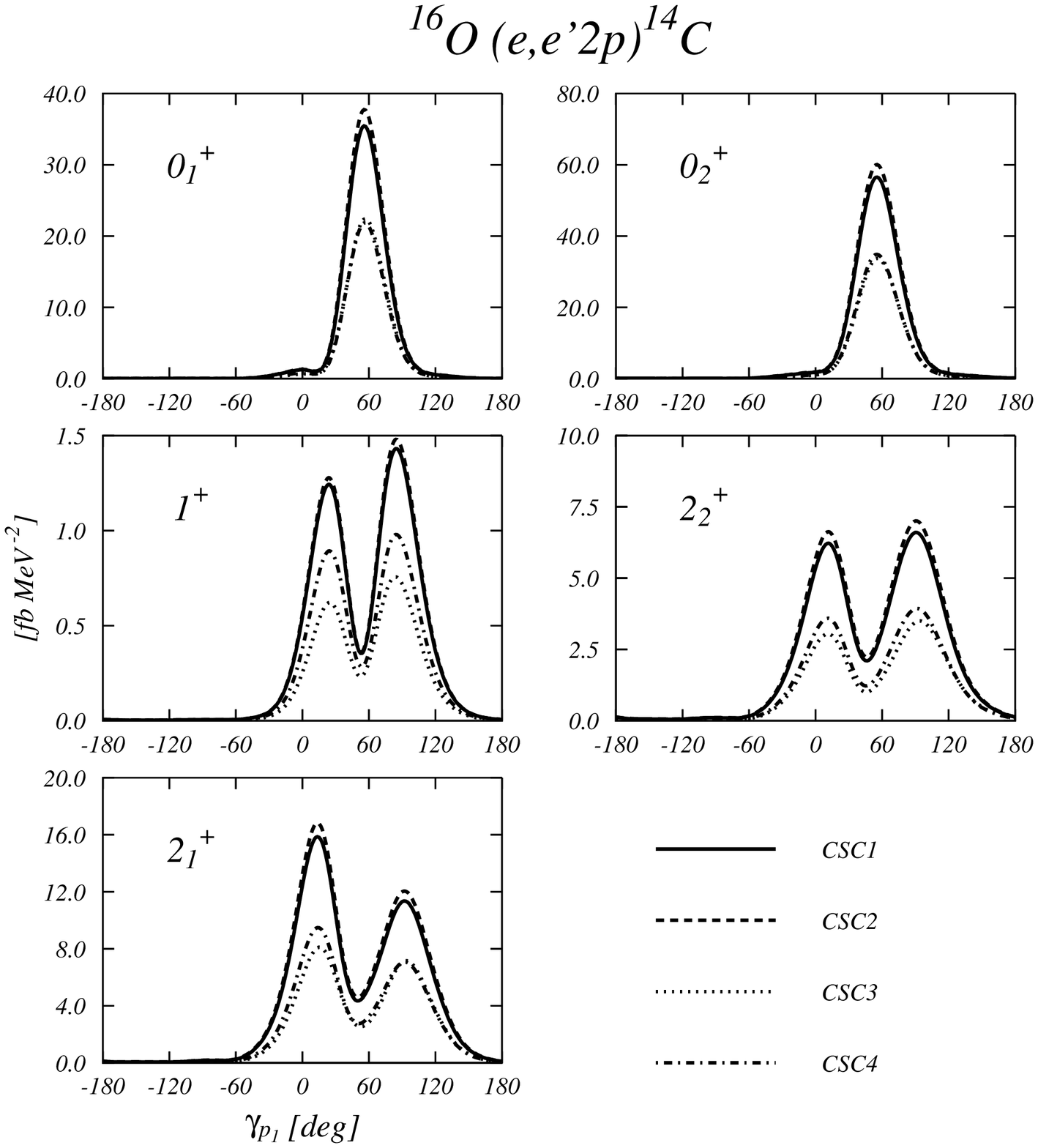}
\vskip 1.0 cm 
\caption{\small Cross section in {\sl standard kinematics} calculated
       with the schematic correlations of Eq. (\protect\ref{eq:scat})
       using the parameters given Table \protect\ref{tab:scat}.
}
\label{fig:box}
\end{figure}

The correlation function $CSC1$ is a square well of radius 1.0 fm and
depth 0.2. The $^{16}$O(e,e'2p)$^{14}$C cross sections calculated in
{\sl standard kinematics} with this correlation function are shown in
Fig. \ref{fig:box} by the full lines. We shall discuss the differences
produced by changing the correlations with respect to these results.
The correlation function $CSC2$ has the same shape of $CSC1$ but a
lower minimum. The effect is a small increase of the cross sections
(see dashed curves). The $CSC3$ correlation function has a square
shape of the type of that of $CSC1$ but, for 1.0 fm$\le r \le$ 1.3 fm,
it overshoots the asymptotic value of 1.0. This modification in the
correlation function is enough to lower to half, roughly, the cross
section maxima (dotted curves). An analogous effect is obtained by the
$CSC4$ which has the same shape as $CSC1$ but a smaller healing
radius. The results obtained with this last correlation function are
shown by the dashed-dotted curves.

The behaviors just described can be understood by remembering that
the quantity entering in the cross section calculation is not $f(r)$
but rather $h(r)=f^2(r)-1$. The largest is the contribution of $h(r)$
in Eq. (\ref{eq:ximod2}), the largest are the cross sections. The
$CSC2$ correlation has a larger overlap with the other integrated
functions than $CSC1$, since its depth is deeper. The overshooting in
$CSC3$ generates a term in $h(r)$ of opposite sign with respect to the
rest of the function, therefore the total contribution to the integral
becomes smaller. The same effect can be obtained by reducing the range
of the correlation, as it is done in $CSC4$.

It is now easier to understand the results of Figs.  \ref{fig:c12},
\ref{fig:s3v8} and \ref{fig:ca40}.  The three correlations used in the
calculations reach their asymptotic values at about the same
internucleon distance: 2 fm. The $S3$ and $V8$ correlation functions
overshoot this value in the region between 1 fm and 2 fm. This part
has opposite effects as the part below the value of 1, as it has been
showing the case of $CSC3$. Since these functions should be multiplied
by a $r^2$ in the integral, the effect of the overshooting is larger
than the small differences in the minima at $r=0$ fm.

It is remarkable that very schematic correlation functions as those used
in Fig. \ref{fig:box} produce angular distributions with the same form
as those of Fig. \ref{fig:s3v8}, obtained with realistic correlation
functions.  This confirms again that the shapes of the angular
distributions are determined by the quantum numbers of the final
states.

\begin{figure}[ht]
\hspace*{2.5cm}
\includegraphics[bb=50 200 500 700,angle=0,scale=0.7]
       {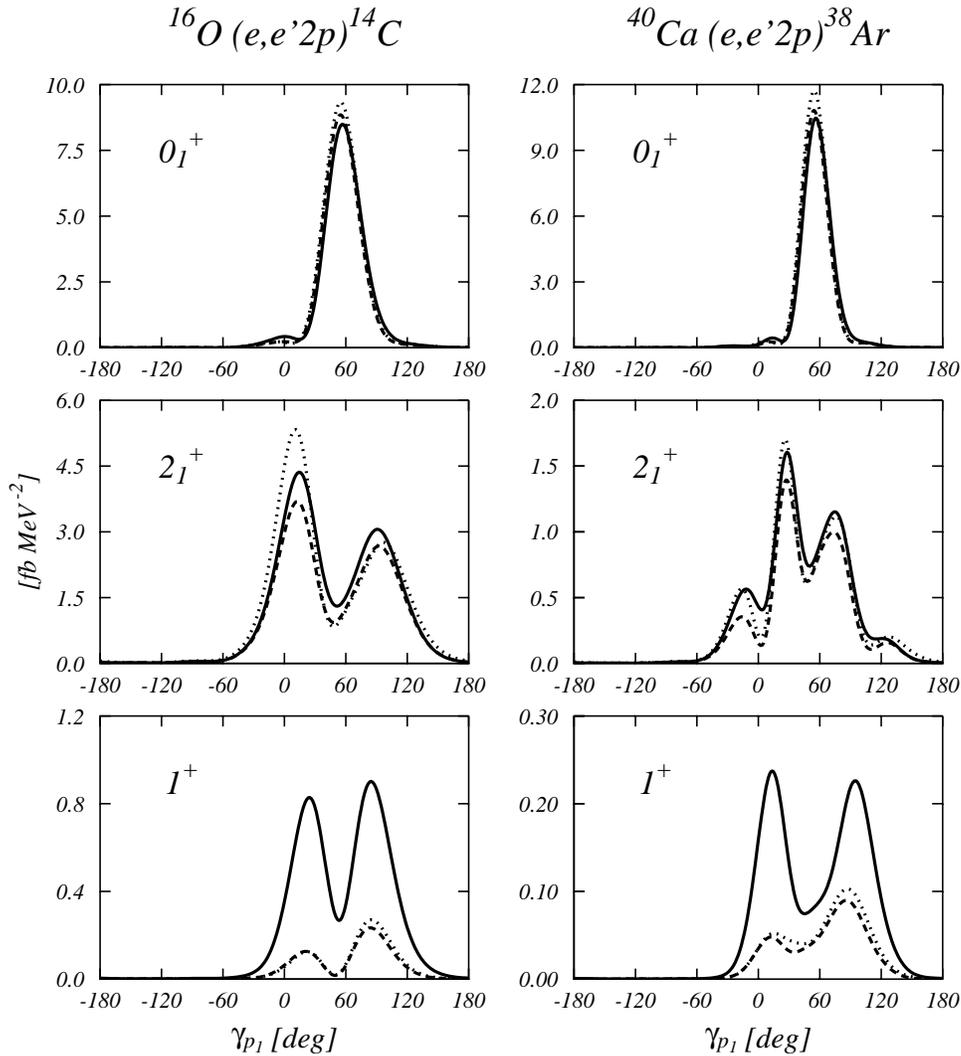} 
\vskip 1.0 cm 
\caption{\small Two nucleon emission cross sections calculated in
       {\sl standard kinematics}. The dotted lines have been obtained
       considering only the 2-point diagrams of
       Fig. \protect\ref{fig:meyer}. The dashed lines have been
       obtained by adding the 3-point diagrams and the full lines show
       the results when also the $\Delta$ currents are included. The
       gaussian correlation of Fig. \protect\ref{fig:corr} has been
       used. 
}
\label{fig:2point}
\end{figure}

\begin{figure}[ht]
\hspace{2.5cm}
\includegraphics[bb=50 200 500 700,angle=0,scale=0.7]
       {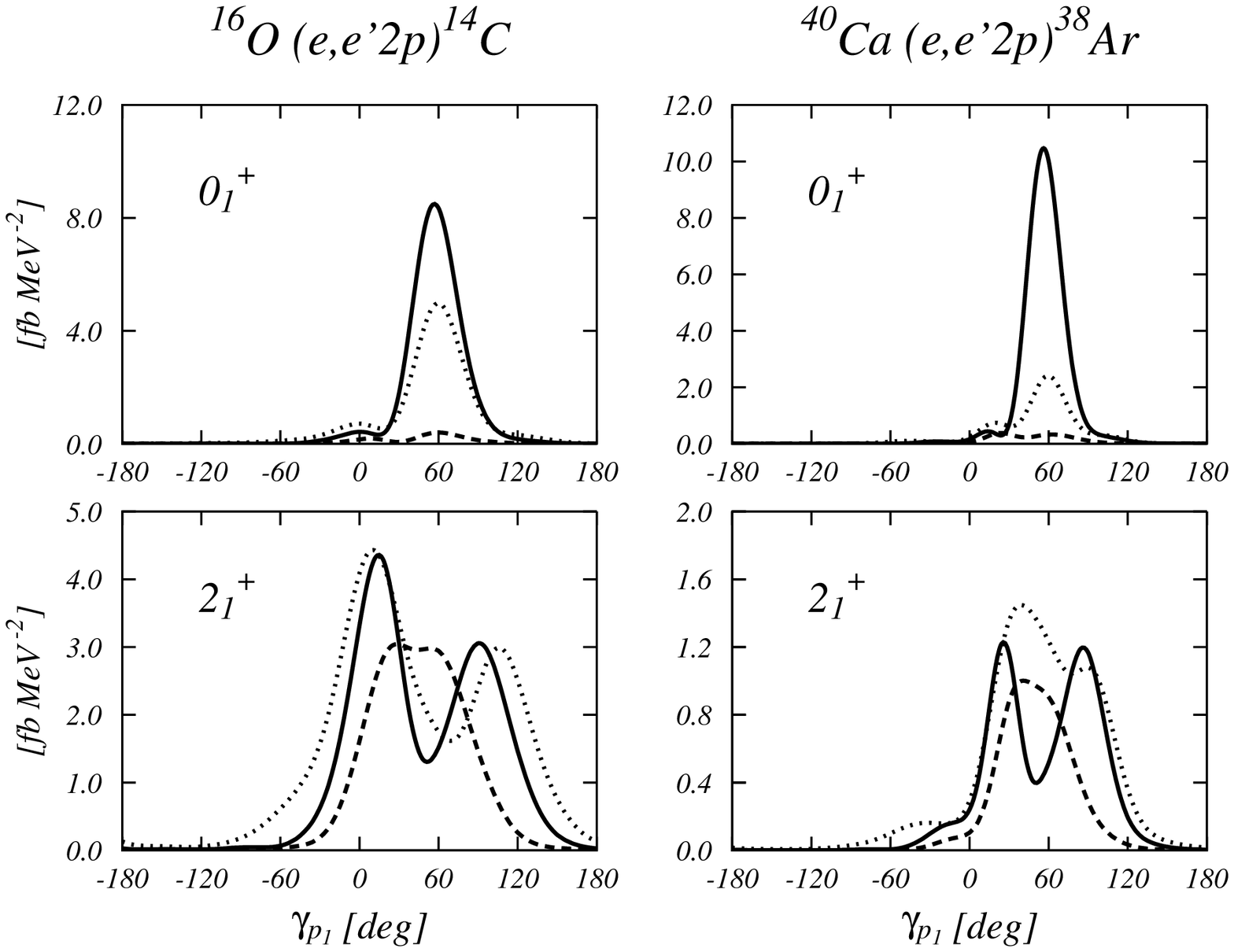} 
\vskip -3.0 cm 
\caption{\small Cross sections calculated in the {\sl standard
       kinematics} but changing the value of the $\gamma_{\ptw}$
     angle. Dotted lines  $\gamma_{\ptw}$=30$^0$, full lines
     $\gamma_{\ptw}$=60$^0$ and dashed lines  
     $\gamma_{\ptw}$=90$^0$. 
}
\label{fig:g2}
\end{figure}

We have tested the role of various diagrams considered in our model.
In Fig. \ref{fig:2point} the cross sections calculated in the {\sl
standard kinematics} and with the gaussian correlation function are
presented. The dotted lines have been obtained by using the 2-point
diagrams of Fig. \ref{fig:meyer}. The dashed lines show the results we
got when 2 and 3-point diagrams of Fig. \ref{fig:meyer} are used. The
total results when also the $\Delta$ current contributions are added
are presented by the full lines.

The inclusion of the 3-point diagrams lowers the cross sections
calculated with the 2-point diagrams only in all the cases we have
considered. This reduction is not constant but it depends from
$\gamma_{\pon}$, as it is clearly shown by the two 2$^+$ states where the
peaks around $\gamma_{\pon}$=0 are more quenched than those around
$\gamma_{\pon}$=100$^0$. The effect of the $\Delta$ currents does not have
a regular behavior. The global quenching of the cross section shown
in $0_1^+$ is followed by and enhancement in the $0_2^+$ state not
shown in the figure.  Also the two $2^+$ states show different
behavior of the $\Delta$ currents. As already mentioned the $1^+$ is
dominated by the $\Delta$ currents even at 100 MeV.

We have already mentioned that $\bpi_r = \bqu - \bpi_1 -
\bpi_2$ is a crucial quantity in the two nucleon emission process.
We have changed this quantity in our {\sl standard kinematics} by
modifying the $\gamma_{\ptw}$ angle. In Fig. \ref{fig:g2} we present
some of the results we have obtained using $\gamma_{\ptw}$=30$^0$ (dotted
lines), $\gamma_{\ptw}$=60$^0$ (full lines) and $\gamma_{\ptw}$=90$^0$ (dashed
lines).

It is interesting to notice that the angular distributions of the
$0^+$ states do not change their form by changing $\gamma_{\ptw}$,
only the size of the peak is modified. The situation is rather
different for the $2^+$ states. By changing $\gamma_{\ptw}$ there is a
merging of the two peaks which are well separated at
$\gamma_{\ptw}$=60$^0$. This situation is analogous to that discussed
in the description of Fig. \ref{fig:resp}.

\section{SUMMARY AND CONCLUSIONS}
\label{sec:con}
We have studied electron induced two-nucleon emission processes by
applying a model which has been already used to describe inclusive
\cite{co01,mok00} and one nucleon emission processes
\cite{mok01,ang02}.  In this model, inspired to the correlated basis
function theory, the full cluster expansion is truncated to consider
all, and only, those diagrams containing a single correlation function
$h(r)$.

In addition to the traditional two-point diagrams, evaluated also in
other approaches \cite{giu97,ryc97}, we consider also three-point
diagrams.  These diagrams are necessary to conserve the proper
normalization of the nuclear final state. The presence of these
tree-point diagrams always reduces the two-point cross sections.  From
the quantitative point of view, in the kinematics we have
investigated, we found relevant effects only for the $2^+$ final
states, where we could detect differences of about 40\%. In all the
other cases the role of the three-point diagrams is much smaller.

We have shown two-proton emission cross sections as a function of the
angle of one of the emitted protons. Because of the energy and momentum
conservation, Eqs. (\ref{eq:econ}) and (\ref{eq:qcon}), this kinematic set
up implies the change of the energy of the proton for every emission
angle. We have used an approximation consisting in neglecting the
recoil of the $A-2$ residual nucleus, therefore the energy of the
emitted nucleons is the same for every emission angle. 
This approximation works well in
the region where the cross sections have their maxima. This indicates
that these maxima appears when the recoil momentum of the residual
nucleus has its minimum value, as was already discussed in
\cite{giu97,ryc96,ryc97}. Since the cross sections values drop by
several order of magnitude in the regions outside the maximum, we used
this no-recoil approximation in our study.

We have investigated the sensitivity of our results to the possible
changes of the mean-field input. We have modified the mean field
parameters such as the description of the nuclear data different from
(e,e'2p) reaction would not be compromised. For example, the hole wave
functions have been modified but we always cared to reproduce the
charge root mean square radii. We used different optical potentials
taken from the literature but all of them reproduce elastic scattering
proton-nucleus data in different energy regimes. We have estimated
that plausible changes of the mean-field bases would imply a theoretical
uncertainty of about 20\% - 30\% on the maxima of the (e,e'2p) cross
section.

The cross sections are most sensitive to the changes in the particle
wave functions. Calculations done considering these wave functions to
be plane waves, or by using purely real Woods-Saxon waves, show cross
sections a factor two larger than those obtained with a complex optical
potential. A consistent description of the one nucleon emission
processes \cite{mok01,ang02} requires the use of complex optical
potentials to take into account the excitation of emission channels
different from the one considered. 
The effects of changing optical potential are certainly smaller than
those produced by the use of real mean-field, but they are not
irrelevant. In the kinematics we have studied we found changes as
large as 18\%. 

The presence of the two-body $\Delta$ currents interferes with the SRC
in the two-nucleon emission mechanism. To minimize the effects of the
$\Delta$ currents one has to find proper kinematics conditions. We
found that the role of these two-body currents at excitation energy
$\omega$=200 MeV is always relevant for every final state. This is
reasonable since 200 MeV is well inside the $\Delta$ resonance region.

The situation changes for $\omega$=100 MeV. The $0^+$ states leaving
the residual nuclei in their ground states are rather insensitive to
the $\Delta$ currents for $|\bqu| \leq $ 500 MeV/c, since their
excitation is mainly of longitudinal type. The $1^+$ states are
instead excited via transverse responses and, even at 100 MeV, they
are strongly affected by the presence of the $\Delta$. The situation
is intermediate for the $2^+$ states where both longitudinal and
transverse excitations are important.

The test of sensitivity of the cross section on the details of the
correlation function has been done by using three different correlation
functions taken from Fermi hypernetted chain calculations on finite
nuclear systems \cite{ari96,fab00}. These are realistic correlation
functions and differ only for few details.

A first remark about our results is that the shape of the angular
distributions of the cross sections remain essentially unchanged. 
This shape is insensitive to the correlations and it is
determined by the decaying waves allowed by the angular momentum and
parity composition related to the quantum number of the nuclear final
state. We essentially confirm the arguments of Ref. \cite{giu98} where
a partial wave decomposition of the various nuclear final states is
presented. 

We have already mentioned that the maxima appear to emission angles
corresponding to the minimum values of the nuclear recoil momentum
$\bpi_r$. The modification of the shapes due to the changes of the
kinematics can be understood in terms of Fourier transform of the
two-hole relative wave function with respect to $\bpi_r$. 

The SRC act on the size of the cross section. Calculations done with a
set of schematic correlation functions, allowed us to understand the
observed behavior in the results obtained with the realistic
correlation functions. The size of the cross section is related to the
function $h(r)=f^2(r)-1$ where $f(r)$ is the two-body correlation
function as it is usually defined. The cross section is more sensitive
to the length of the healing distance, the distance where $h(r)$
becomes zero, than to the value of $h(r)$ for $r=0$. An overshooting
of the asymptotic value of $f(r)$, corresponding to a change of sign
of $h(r)$, produces results similar to those obtained by reducing the
length of the healing distance.

The information about the SRC can be obtained only by a quantitative
comparison between theoretical predictions and experimental data.
Qualitative features, such as the shape of the angular distributions,
are not sensitive to the details of the SRC. Unfortunately a precise
quantitative evaluation of the (e,e'2p) cross sections is linked to
the theoretical framework used to calculate it, and to the
uncertainties in the required theoretical input. It appears clear that
two-nucleon knock out experiments cannot be considered as the ultimate
tool able to determine the characteristics of the SRC correlations.
Instead, they have to be considered as another, useful and interesting,
element of a puzzle, that together with elastic, inclusive and
one-nucleon emission experiments should be described in a unique
and coherent theoretical framework.

%
%
\section*
{APPENDIX A: Matrix elements of the one-body operators}
In this appendix we give the explicit expressions of the matrix
elements (\ref{eq:psif2}) needed to evaluate the response functions of
Eqs. (\ref{eq:wl})-(\ref{eq:wtt}). Actually, what we calculate
is the $\xi^1$ term as defined in Eq. (\ref{eq:ximod2}). The
calculation of the various terms defining this transition matrix
element proceeds by making a multipole expansion of the
correlation function $h_{ij}$:
\begin{equation}
h_{ij} \,= \,h(r_{ij}) \,= \, h(r_i,r_j,\cos \theta_{ij}) \,
 =  \, \sum_{L=0}^\infty  \, h_L(r_i,r_j)  \, P_L(\cos \theta_{ij})
\,\, ,
\label{eq:hexp}
\end{equation}
where $P_L$ are the associated Legendre polynomials. 

As shown in Eq. (\ref{eq:opft}), the operator
$O_{\eta}(\bqu)$ is the Fourier transform of a one-body operator, in our case the
electromagnetic operators (\ref{eq:charge1}) and (\ref{eq:mag}). We perform
a multipole expansion of the $e^{- i \bqu \cdot \br }$ term and,
because we have chosen the $z$-axis parallel to the direction of the
momentum transfer $\bqu$ we obtain
\begin{eqnarray}
\label{eq:charge}
O_0(\bqu) \equiv \rho (\bqu) &=&
\sqrt{ 4 \pi} \, 
\sum_J \, (-i)^J \, \what{J} \, \int {\rm d}^3 r \, j_J(qr) \, Y_{J0}(\Omega)
\, \rho (\br)  \, , \\
\nonumber
\label{eq:current}
O_{\pm 1}(\bqu) \equiv J_{\pm 1} (\bqu) &=& 
\sqrt{ 2 \pi} \, \sum_J \, (-i)^J \, \what{J} \, \int {\rm d}^3 r 
\left[
 (-i)^{-1} \, \frac{\sqrt{J+1}}{\what{J}} \, j_{J-1}(qr) \,
{\bf Y}^{\pm1}_{J-1,1,J}(\Omega)
\right.\\
&~& \left.
\pm \, j_J(qr) \, {\bf Y}^{\pm1}_{J,1,0}(\Omega) \,
+ \, (-i) \, \frac{\sqrt{J}}{\what{J}} \, j_{J+1}(qr) \,
{\bf Y}^{\pm1}_{J+1,1,J}(\Omega) 
\right] \, \cdot \, J(\br)
\,\, ,
\end{eqnarray}
where we have indicated with
\begin{equation}
{\bf Y}^M_{\lambda,1,J}(\Omega) \, = \, \sum_{\mu \eta} 
<\lambda \mu 1 \eta | J  M > Y_{\lambda \mu}(\Omega)
\, e_\eta
\end{equation}
the vector spherical harmonics \cite{edm57} and with $j_\lambda$ the
spherical Bessel functions.  In the above equations we have defined
$\what{l} = \sqrt{2l+1}$.

From the above equations and the expressions
(\ref{eq:charge1}) and (\ref{eq:mag}) we found convenient to express
each multipole component of the one-body operator operator $O_\eta$ in terms 
of products of a factor depending on $qr$ and another one depending only on
the angular coordinates $\Omega$. For the charge operator ($\eta = 0$)
we have
\begin{equation}
O_0(\bqu) \, = \, \sum_J \int {\rm d}^3r \,
F_J(qr)\, {\cal O}_{J,0}(\Omega) 
\, .
\label{eq:genob}
\end{equation}
By taking into account Eqs. (\ref{eq:charge1}) and (\ref{eq:charge}) we
identify
\beq
F_J(qr) \, = \, \sqrt{ 4 \pi} \, (-i)^J \what{J}  j_J(qr) \sum_{i=1}^A \,
\left[ G^{\rm P}_{\rm E}(\bqu) \, \frac{1+\tau_3(i)}{2}
+      G^{\rm N}_{\rm E}(\bqu) \, \frac{1-\tau_3(i)}{2}
\right]
\, ,
\eeq
and 
\beq
{\cal O}_{J,0}(\Omega) \, = \, Y_{J0}(\Omega)
\, .
\eeq
In the above equation we have indicated with $G^{\rm P,N}_{\rm E}(\bqu)$ the
electric form factor of the proton and neutron, respectively. 

For the magnetization current ($\eta = \pm 1$) we have
\begin{equation}
O_{\pm 1}(\bqu) \, = \, \sum_J \, \int {\rm d}^3r \, \sum_{s=\pm1,0} \,
F^s_J(qr) \, {\cal O}^s_{J,\pm 1}(\Omega) \, .
\end{equation}
Taking into account Eqs. (\ref{eq:mag}) and (\ref{eq:current}) 
the factors in the previous equation are defined as:
\beq
F^s_J(qr) \, = \,
\sqrt{ 2 \pi} \, (-i)^J \, \Gamma(J,s) \, j_{J+s}(qr) \, \sum_{i=1}^{A} \,
\left[ \frac{G^{\rm P}_{\rm M}(\bqu)}{2 m_i} \, \frac{1+\tau_3(i)}{2}
+      \frac{G^{\rm N}_{\rm M}(\bqu)}{2 m_i} \, \frac{1-\tau_3(i)}{2}
\right]
\, ,
\eeq
with $G^{\rm P,N}_{\rm M}(\bqu)$ the nucleon magnetic form factor and
\beq
\Gamma(J,s) \, = \, \left\{ \begin{array}{ll}
                      - i \, \eta \, \sqrt{J + \delta_{s,-1}} \, , & s=-1 \\
                           \what{J} \, ,                     & s= 0 \\
                        i \, \eta \, \sqrt{J} \, ,                 & s=+1 \\
                      \end{array}
               \right.
                      \, ,
\eeq
and
\beq
{\cal O}^s_{J,\eta}(\Omega) \, = \, \left [Y_{J+s}(\Omega) \otimes \bsigma
\right]^J_\eta
\, .
\eeq

In the following, the coordinate where the one-body operators is
acting will be labeled $r_1$. The same notation has been used in Eq.
(\ref{eq:ximod2}).
For the charge we define the integrals
\beq
K(q;J;a,b) \equiv 
\int {\rm d} \ron \, \ron^2 \, 
F_J(q\ron) \, R^*_a(\ron) \, R_b(\ron)
\, ,
\label{eq:KJ}
\eeq
\beq
I(q;J,L;a,b,c,d) \equiv 
\int {\rm d} \ron \, \ron^2 \, \int {\rm d} \rtw \, \rtw^2 \, h_L(\ron,\rtw) 
\, F_J(q\ron) \, R^*_a(\ron) \, R^*_b(\rtw) \, R_c(\ron) \, R_d(\rtw)
\, ,
\label{eq:IJL}
\eeq
and
\beq
H(L;a,b,c,d) \equiv 
\int {\rm d} \ron \, \ron^2 \, \int {\rm d} \rtw \, \rtw^2 
\, h_L(\ron,\rtw) \,
R^*_a(\ron) \, R^*_b(\rtw) \, R_c(\ron) \, R_d(\rtw)
\, ,
\label{eq:HL}
\eeq
where $R$ is the radial part of the single particle wave function 
defined in Eq. (\ref{eq:spwf}). Similar expressions are defined
for the current operator by changing $F_J(q\ron)$ into $F^s_J(q\ron)$
in the $K$ and $I$ integrals.

The orthonormality of the single particle basis used to construct the
Slater determinants $|\Phi_{\rm f}>$ and $|\Phi(0,0)>$ ensures that
only the single particle wave functions directly involved by the
electromagnetic operator and by the correlation function contribute to
the matrix elements.

Following the nomenclature of the diagrams shown
in Fig. \ref{fig:meyer} we obtain
\begin{eqnarray}
\nonumber
\langle O \rangle _{(2.1)} \, \equiv & &  
\sum_{JL\lambda} \, I(q;J,L;\pon,\ptw,\hon,\htw) \,
\delta_{t_{\pon},t_{\hon}} \, \delta_{t_{\ptw},t_{\htw}} \, 
\frac {4 \pi} {\what{L}^2} \, \what{\lambda} \,
(-1)^{ J-L+\eta+j_{\pon}-m_{\pon}+j_{\ptw}-m_{\ptw} }\\
\nonumber
&~&
 \, 
\threej{J}{L}{\lambda}{\eta}{-N}{N-\eta} \, 
\threej{j_{\pon}}  {\lambda} {j_{\hon} }
       {-m_{\pon}} {-N+\eta} { m_{\hon}} \, 
       \threej{j_{\ptw}}  {L} {j_{\htw} }
       {-m_{\ptw}} {N} { m_{\htw}} \\
&~&
\langle j_{\pon} || 
\left[ {\cal O}_J  \otimes Y_L \right]^\lambda 
|| j_{\hon} \rangle 
\langle j_{\ptw} || Y_L || j_{\htw} \rangle 
\, ,
\label{eq:g21}
\end{eqnarray}
where $\delta_{a,b}$ is the Kronecker symbol. The expression of the (2.2)
diagram is obtained from the above one by interchanging $\hon$ and
$\htw$. The expressions of the (2.3) and (2.4) diagrams are obtained
by interchanging $\pon$ with $\ptw$ in the
expressions of the (2.1) and (2.2) diagrams, respectively.

For the three-point diagrams we have
\begin{eqnarray}
\nonumber
\langle O \rangle _{3.1} \equiv & &
\sum_{JL} \, \sum_i \, 
K(q;J;\pon,i) \, H(L;i,\ptw,\hon,\htw) \,
\delta_{t_{\pon},t_i} \, \delta_{t_i,t_{\hon}} \, 
\delta_{t_{\ptw},t_{\htw}} \\ 
\nonumber
&&
\frac {4 \pi} {\what{L}^2} \, 
(-1)^{ j_{\pon}-m_{\pon} } \, (-1)^{ j_{\ptw} - m_{\ptw} } \, 
(-1)^{ N + \ji - \mi } \\
\nonumber
&~&
\threej{ j_{\pon}}  { J } {\ji }
       {-m_{\pon}}  {\eta} { \mi }
\,
\threej{ j_{\ptw}  }  {  L }  { j_{\htw} }
       {-m_{\ptw}  }  {  N } { m_{\htw} }
\,
\threej{ \ji }  {  L }  { j_{\hon} }
       {-\mi }  {- N } { m_{\hon} }
 \\
&~&
\langle j_{\pon} || {\cal O}_J || \ji \rangle 
\,
\langle  j_{\ptw}  || Y_L || j_{\htw} \rangle 
\,
\langle \ji || Y_L || j_{\hon} \rangle
\, .
\label{eq:g31}
\end{eqnarray}
The expression of the diagram (3.2) is analogous to the above
one but interchanging $\hon$ with $\htw$.  The (3.3) and
(3.4) diagrams are obtained by interchanging $\pon$ and $\ptw$
 in the (3.1) and (3.2) diagrams respectively. For
the other diagrams we have:
\begin{eqnarray}
\nonumber
\langle O \rangle _{3.5} \equiv &~&
\sum_{JL} \, \sum_i \, 
K(q;J;i,\htw) \, H(L;\ptw,\pon,i,\hon) \,
\delta_{t_i,t_{\htw}} \, \delta_{t_{\ptw},t_i} \, 
\delta_{t_{\pon},t_{\hon}} \\
& &
\nonumber
\frac {4 \pi} {\what{L}^2} \, (-1)^{\ji - \mi } 
\, (-1)^{ j_{\pon}-m_{\pon} } 
\, (-1)^{ N + j_{\ptw} - m_{\ptw} } 
\\
\nonumber
&~&
\threej{ \ji }  { J } {j_{\htw} }
       {-\mi}  {\eta} {m_{\htw} }
\, \threej{ j_{\pon}  }  {  L }  { j_{\hon} }
       {-m_{\pon}  }  {  N } { m_{\hon} }
\, \threej{ j_{\ptw} }  {  L } { \ji }
       {-m_{\ptw} }  {- N } { \mi }
 \\
&~&
\langle \ji || {\cal O}_J || j_{\htw} \rangle 
\, \langle  j_{\pon}  || Y_L || j_{\hon} \rangle 
\, \langle j_{\ptw} || Y_L || \ji \rangle
\, .
\label{eq:g35}
\end{eqnarray}
Also in this case we obtain the expression of the (3.6) diagram 
from the above expression by interchanging $\hon$ and $\htw$. 

The above expressions are specialized for the charge operator. For the
magnetization current, ${\cal O}_J$ must be changed into ${\cal
  O}^s_{J,\eta}$.  The various charge and current transitions are
calculated by substituting the above definitions in Eqs.
(\ref{eq:g21})-(\ref{eq:g35}) and remembering that
\begin{eqnarray}
<\ja \| Y_L \| \jb> & = &
(-1)^{\ja+\half} \, \frac{\what {\ja} \what {\jb} \what {L}}{\sqrt{4\pi}} \,
\threej {\ja}{L}{\jb}{\half}{0}{-\half} \, \xi(\la+\lb+L)  
\, , \\
<j_a \|[Y_L \otimes Y_J ]^K \| j_b> &=& (-1)^{L-J} \,
\frac{\what{J} \what{L} }{\sqrt{4\pi}} \,
\threej{L}{K}{J}{0}{0}{0} \,
<j_a \|Y_K \| j_b>
\, , \\
<j_a \|[Y_J \otimes \bsigma ]^J \| j_b> & = & (-1)^\la \,
\frac{ \what{\ja} \what{\jb} \what{J}} {\sqrt{4\pi}} \,
\threej{\jb}{\ja}{J}{-\half}{-\half}{1}  \, \xi(\la + \lb + J)
\, , \\
\nonumber
<j_a \|[Y_{J_s} \otimes \bsigma ]^J \| j_b> &=& 
(-1)^{\la+\lb+\jb+\half} \,
\frac{ \what{\ja} \what{\jb}} {\sqrt{4\pi}} \,
\frac{\chi_a + \chi_b + sJ + \delta_{s,1}}
{\sqrt{J+\delta_{s,1}}} \\
& &
\threej{\ja}{\jb}{J}{\half}{-\half}{0} \, \xi(\la + \lb + J +1)
\, ,
\end{eqnarray}
where $|\ja > \equiv |\lb \half \ja>$, with $\xi(L)=1$ if $L$ is even and zero otherwise 
and
\beq
\chi_a \equiv (-1)^{\la+\ja+\half}(\ja+\half)=(\la-\ja)(2\ja+1)
\, .
\eeq
\vskip 2.cm 
\section*{ACKNOWLEDGMENTS}
We thank Carlotta Giusti for her interest in our work, the numerous
discussions and for the careful reading of the manuscript.
We also thank Paolo Christillin and  Roberto Perrino for
useful discussions. This work has been partially supported by the
agreement INFN-CICYT, by the DGES (PB98-1367), by the Junta de
Andalucia (FQM 225) and by the MIUR through the PRIN {\sl Fisica del
  nucleo atomico e dei sistemi a molticorpi}.
%
%

%
%
%
\newpage
%
%
%
%
%
%
%
%
%
%
%
%
\newpage
%
%
%
%
%
%
\clearpage

\end{document}